\documentclass[11pt]{article}
\usepackage{jheppub}
\usepackage{diagbox}
\usepackage{amsmath}
\usepackage{graphicx}
\usepackage{epstopdf}
\usepackage{forest}	
\usepackage{subfigure}
\usepackage{multirow}
\usepackage{caption}
\usepackage{amsfonts,amssymb}
\usepackage[numbers,sort&compress]{natbib}
\usepackage{makecell}
\usepackage{dsfont,bbm}
\usepackage{youngtab}
\usepackage{arydshln}
\usepackage{bm}

\allowdisplaybreaks[4]
\graphicspath{{./figs/}}

\newcommand\symbolOa{\mathcal{O}}

\title{Gluonic evanescent operators: negative-norm states and complex anomalous dimensions}
\author[b]{Qingjun Jin,}
\emailAdd{qjin@gscaep.ac.cn}
\author[a]{Ke Ren,}
\emailAdd{renke@itp.ac.cn}
\author[c,d,e]{Gang Yang,}
\emailAdd{yangg@itp.ac.cn}
\author[f,g]{and Rui Yu}
\emailAdd{yurui@imu.edu.cn}
\affiliation[a]{School of Physics and Astronomy, Sun Yat-Sen University, Zhuhai 519082, China}
\affiliation[b]{Graduate School of China Academy of Engineering Physics, \\ No.~10 Xibeiwang East Road, Haidian District, Beijing, 100193, China}
\affiliation[c]{CAS Key Laboratory of Theoretical Physics, Institute of Theoretical Physics, \\Chinese Academy of Sciences, Beijing 100190, China}
\affiliation[d]{School of Fundamental Physics and Mathematical Sciences, Hangzhou Institute for Advanced Study, UCAS, Hangzhou 310024, China}
\affiliation[e]{International Centre for Theoretical Physics Asia-Pacific, Beijing/Hangzhou, China}
\affiliation[f]{School of Physical Science and Technology, Inner Mongolia University, Hohhot 010021, China}
\affiliation[g]{Beijing Computational Science Research Center, Beijing 100193, China}

\abstract{
In this paper, we build on our previous work to further investigate the role of evanescent operators in gauge theories, with a particular focus on their contribution to violations of unitarity. 
We develop an efficient method for calculating the norms of gauge-invariant operators in Yang-Mills (YM) theory by employing on-shell form factors. Our analysis, applicable to general spacetime dimensions, reveals the existence of negative-norm states among evanescent operators. We also explore the one-loop anomalous dimensions of these operators and find complex anomalous dimensions. 
We broaden our analysis by considering YM theory coupled with scalar fields and we observe similar patterns of non-unitarity. 
The presence of negative-norm states and complex anomalous dimensions across these analyses provides compelling evidence that general gauge theories are non-unitary in non-integer spacetime dimensions. }

\begin{document}
\maketitle

\section{Introduction}

Evanescent operators are a unique category of operators that, while vanishing in four space-time dimensions, remain non-zero in general non-integer dimensions. The four-fermion type evanescent operators have been studied a long time ago, see \emph{e.g.}~\cite{Buras:1989xd,Dugan:1990df,Herrlich:1994kh,Buras:1998raa}.
The systematic classification and two-loop renormalization effects of gluonic evanescent operators in Yang-Mills (YM) theory have been recently addressed in \cite{Jin:2022ivc, Jin:2022qjc}.  
In this paper, we delve into a remarkable property of evanescent operators in gauge theories: 
their role in revealing the  non-unitary nature of gauge theories in non-integer spacetime dimensions.

There are two principal motivations for studying Quantum Field Theory (QFT) in non-integer dimensions. The first arises from the dimensional regularization technique, which analytically continues spacetime from four to $d=4-2\epsilon$ dimension to manage ultraviolet or infrared divergences, such as those encountered in loop integrals \cite{tHooft:1972tcz}.
The second is rooted in the $\epsilon$-expansion method employed to calculate critical exponents \cite{Wilson:1971dc} where the analytic continuation serves to bridge theories across various integer dimensions. 
In particular, theories at the conformal fixed points are defined in general non-integer dimensions depending on $\epsilon$.
 
Unitarity, the bedrock principle ensuring probability conservation in QFT, has been challenged in non-integer spacetime dimensions. 
This was first suggested in the study of scalar $\phi^4$ theory by the existence of 
negative-norm states which are related to the evanescent operators,
and the presence of complex anomalous dimensions (ADs) at the Wilson-Fisher (WF) fixed point  \cite{Hogervorst:2014rta, Hogervorst:2015akt}.
In this paper, we show that the gluonic evanescent operators in YM theory exhibit similar characteristics, suggesting that unitarity violation is
a ubiquitous feature in QFT at non-integer spacetime dimensions.
Part of the results have been reported in a brief letter \cite{Jin:2023cce}, and in this paper, we provide a comprehensive exposition of the computational techniques and a detailed presentation of our results.

Our methodology involves analyzing the norm of operators through their two-point functions of gauge-invariant local operators.
We have developed an efficient approach by working in momentum space and utilizing form factors as building blocks. 
This method is applicable in general $d$ dimensional formalism and thus applies to evanescent operators.
Explicit computations are done for the norm of operators up to $\Delta_0 = 16$, here $\Delta_0$ is the canonical dimension of the operators. 
The evanescent operators, while being null states in four spacetime dimensions, can yield negative-norm states in non-integer spacetime dimensions.

We also examine one-loop anomalous dimensions for high-dimensional gauge-invariant operators. 
The anomalous dimensions are given at the leading order in the $\epsilon$ expansion at the WF fixed point.
Our analysis of Gram matrices indicates that complex anomalous dimensions associated with evanescent operators emerge starting from $\Delta_0=12$, which is verified by the results.
Interestingly, we find the numbers of complex anomalous dimensions match with the number of negative-norm states for all length-4 operators, which is checked up to $\Delta_0=16$.
It is noteworthy that the complex anomalous dimensions appear at a relatively low mass dimension $\Delta_0=12$ in YM theory, contrasting with the scalar $\phi^4$ theory, where they do not appear until  $\Delta_0=23$.

The remainder of this paper is organized as follows.
After a brief review of some basics of the YM theory operators in Section~\ref{sec:setup}, we study the Gram matrix of operators in Section~\ref{sec:GramMatrix}.
The method of computing the Gram matrix based on form factors is introduced, and the results of Gram matrices for various operator sectors are presented, where the appearance of negative-norm states is discussed.
In Section~\ref{sec:complexAD}, we study the one-loop anomalous dimensions.
Specifically, we see the appearance of complex anomalous dimensions signifying the violation of unitarity for both the pure YM theory and a generalized gauge theory with YM coupled to scalars.
We conclude with a summary and discussion in Section~\ref{sec:discuss}.
Appendices provide technical details and explicit results, while ancillary files offer a comprehensive repository of operator bases, Gram matrices, and anomalous dimensions.

\section{Set up}
\label{sec:setup}

The operator types discussed in this paper are gauge-invariant Lorentz scalar operators in
pure Yang-Mills theory.
To facilitate the forthcoming discussion, we will clarify the operators and convention from the outset.

The Yang-Mills operators are composed of field strength $F_{\mu\nu}$ and covariant derivatives $D_\mu$.
The field strength carries a color index as $F_{\mu\nu}=F_{\mu\nu}^a T^a$, where
$T^a$ are generators of gauge group satisfying $[T^a,T^b]=\mathbbm{i}f^{abc}T^c$.
The gauge invariant  Yang-Mills operators have a general form
\begin{equation}
\label{eq:operator}
\mathcal{O}(x)\sim c(a_1, \ldots ,a_L) {\cal W}_{m_1}^{a_1} {\cal W}_{m_2}^{a_2} \ldots {\cal W}_{m_L}^{a_L} \,, \qquad
\textrm{with} \ \ 
{\cal W}_{m_i} = D_{\mu_{i_1}}...D_{\mu_{i_{m_i}}}F_{\nu_i \rho_i}
 \,,
\end{equation}
where the covariant derivative acts as
\begin{align}
\label{eq:derivative}
D_\mu F_{\nu\rho}=\partial_\mu F_{\nu\rho} - \mathbbm{i} g [A_\mu,F_{\nu\rho}]\,,
\end{align}
and $c(a_1,...,a_n)$ is the color factor, and it may be given in terms of traces of products of generators $\mathrm{tr}(...T^{a_i}...T^{a_j}...)$ or as products of structure constants $f^{a_i a_j a_k} ... f^{a_l a_m a_n}$.
In this paper, we consider the Lorentz scalar operators, where
Lorentz indices $\{\mu_i,\nu_i,\rho_i\}$ are contracted pairwisely
among different $\mathcal{W}_{m_i}^{a_i}$.
These operators  are covariant in $d$ dimensional spacetime and $P$-parity even.
We will not consider $P$-parity odd operators involving the Lorentz Levi-Civita tensor. 
We define the \emph{length} of the operator as the number of ${\cal W}_{m_i}^{a_i}$, for example, the operator in \eqref{eq:operator} has length $L$.

In Section~\ref{sec:GramMatrix}, we will also consider operators in the scalar $\phi^4$ theory, which are composed of scalar field $\phi$ and derivative $\partial_\mu$. As illustrated in \eqref{eq:operator}, they can be represented as 
\begin{equation}
\label{eq:scalaroperator}
{\cal O}^\phi(x) \sim \tilde{\cal W}_{m_1} \tilde{\cal W}_{m_2} \ldots \tilde{\cal W}_{m_L} \,, \qquad
\textrm{with} \ \tilde{\cal W}_{m_i} = \partial_{\mu_{i_1}}...\partial_{\mu_{i_{m_i}}} \phi \,,
\end{equation}
where all Lorentz indices are contracted among the derivatives. The length of the operator is given similarly by the number of $\tilde{\cal W}_{m_i}$.
By comparing \eqref{eq:operator} and \eqref{eq:scalaroperator}, it is evident that the YM operators are more complex due to the color degrees of freedom, as well as the fact that field strength tensors carry Lorentz indices.

For notation convenience, we represent Lorentz indices with integer numbers
and abbreviate $D_i D_j\cdots$ as $D_{ij\cdots}$.
The same integers imply that the indices are contracted, such as $F_{12}F_{12}=F_{\mu\nu}F^{\mu\nu}$.
We will explicitly take the gauge group as SU$(N_c)$, although the generalization to other Lie groups is straightforward.

Gauge invariant local operators have an intimate connection with the concept of form factors. These are matrix elements between on-shell states and a local operator ${\cal O}(x)$ as \cite{Maldacena:2010kp,Brandhuber:2010ad, Bork:2010wf, Yang:2019vag}
\begin{align}\label{eq:formfactor}
		\mathcal{F}_{\mathcal{O}, n}=\int \text{d}^dx e^{-{\text{i}}q\cdot x}\langle p_1\cdots, p_n |\mathcal{O}|0\rangle\,,
\end{align}
where $p_i$ denote the on-shell momenta carried by the external gluons, and $q=\sum_i {p_i}$ is the off-shell momentum associated with the operator.
A specific class of form factors is the so-called \emph{minimal form factors} where the number of external on-shell states equals the length of the operator.  
For more details on the form factors of YM operators in $d$-dimensional kinematics, the readers are referred to \cite{Jin:2019opr, Jin:2020pwh, Jin:2022ivc, Jin:2022qjc}.
The link between form factors and local operators makes it possible to apply powerful on-shell amplitude techniques.
As we will demonstrate in the following two sections, form factors play a crucial role in both the computation of the Gram matrices and the renormalization of the operators.

Below, we define several concepts that will be used for operator classification.

\paragraph{Physical and evanescent operators.}

There exist some operators defined at general $d$ dimensions
and  vanishing in the limit $d\rightarrow 4$, which are called \emph{evanescent operators}.
More rigorously, an operator is called
an evanescent operator, if the tree-level matrix elements of this operator have non-trivial
results in general $d$ dimensions but all vanish in four dimensions.
If an operator is
not an evanescent operator, \emph{i.e.} its form factors do not vanish in four dimensions, we call it
a \emph{physical operator}.

An evanescent operator can be constructed by multiplying a tensor operator by a
Kronecker symbol $\delta^{\mu_1\cdots\mu_n}_{\nu_1\cdots\nu_n}$ with rank $n>4$ and then taking Lorentz contraction,
\emph{e.g.}
\begin{align}
\label{eq:deltaeg}
\delta^{1234\mu}_{5678\nu}\, \mathrm{tr}( D_\nu F_{12} F_{34} D_\mu F_{56} F_{78})\,.
\end{align}
A rank-$n$ Kronecker symbol is defined as
\begin{equation}
\label{eq:kronecker1}
\delta^{i_1\cdots i_n}_{j_1\cdots j_n}:= 
\left|
\begin{matrix}
\delta^{i_1}_{j_1} & \ldots & \delta^{i_1}_{j_n} \\
\vdots &  & \vdots\\
\delta^{i_n}_{j_1} & \ldots & \delta^{i_n}_{j_n}
\end{matrix}
\right| \,,
\end{equation}
anti-symmetric among indices lying in the same row and 
therefore vanishes at $d<n$.
If the highest Kronecker symbol contained by an operator is of rank-$n$, 
we say this operator is a $\delta$-$n$ operator, \emph{e.g.}
 (\ref{eq:deltaeg}) is a $\delta$-5 operator.

It is also useful to define the generalized $\delta$-function of rank-$n$
for two lists of Lorentz vectors
$\{u_i\}$, $\{v_i\}$, $i=1,..,n$:
\begin{align}
\label{eq:kronecker2}
\delta^{u_1,...,u_n}_{v_1,...,v_n}&=
\det (u_i\cdot v_j)\,.
\end{align}
The minimal form factor of a $\delta$-$n$ operator
is composed of such $\delta$-functions, where all the 
vectors $u_i$, $v_i$ take values in gluon polarization vectors
and external momenta. 
Considering a $\delta$-$n$ operator with length $L$,
its minimal form factor
is a polynomial of $e_i\cdot p_j,\ e_i\cdot e_j,\ p_i\cdot p_j$ and 
in each term every $e_i$ appears once.
Writing this form factor in the form of $\delta^{e_i p_i\cdots}_{e_j p_j\cdots}$,
two rows together contain $L$ polarization vectors and $2n-L$ momenta.
In each row, every $p_i$ at most appears once otherwise the $\delta$-function becomes zero, 
so one concludes that $2n-L\leq 2L$. In other words,
length-$L$ evanescent operators cannot be constructed
from Kronecker symbols higher than rank $ [3L/2] $.
Correspondingly,   evanescent operators requiring Kronecker
symbols higher than 4 only exist for length $L>3$.

The evanescent operators in scalar   theories can also be constructed
through the Kronecker symbols, see examples in \cite{Hogervorst:2015akt}.
The minimal form factors of scalar evanescent operators depend
on the external momenta only, so they are composed of $\delta$-functions 
$\delta^{p_{i_1}p_{i_2}\cdots}_{p_{j_1}p_{j_2}\cdots}$.
Such rank-$n$ functions require  at least $n$ external momenta to avoid
two same $p_i$ lying in the same row, so
a $\delta$-$n$ operator in scalar theory is at least of length-$n$.
For example, there are no scalar evanescent operators at the level of length-4,
and no $\delta$-6 scalar ones at the level of either length-4
or length-5.
In comparison, both $\delta$-5 and $\delta$-6 operators
are permitted at the level of length$\geq$4 in Yang-Mills theory
since the inequality $2n\leq 3L$ are  satisfied.

\paragraph{Classification of operators.}

The canonical dimension of the operator is determined by the number of covariant derivatives and the field strengths as follows:
\begin{equation}
\Delta_0 = (\# \textrm{ of }D_\mu) + 2 \times(\# \textrm{ of }F_{\mu\nu}) \,. 
\end{equation}
For a given canonical dimension, there are a finite number of operators. However, these operators may not be independent of each other in the sense that
two operators are considered \emph{equivalent} if their difference is proportional to the equation of motion ($D_\mu F^{\mu\nu} = 0$) or the Bianchi identity ($D_\mu F_{\nu\rho}+D_\nu F_{\rho\mu}+D_\rho F_{\mu\nu} = 0$).
In practical terms, it is useful to identify the basis of operators at a given mass dimension by eliminating such equivalence.
The detailed classification of length-4 physical and evanescent operators
with all canonical dimensions can be found in our previous work \cite{Jin:2022ivc}.

At the quantum level, upon renormalization, the operators of the same canonical dimension will generally mix with each other. Consequently, the dimensions of the operators generally receive a correction $\Delta = \Delta_0 + \gamma(g)$, where $\gamma$ is known as the anomalous dimension.
These topics will be covered in more detail in Section~\ref{sec:complexAD}.
To simplify the mixing structure, it is beneficial to classify them further according to their properties so that the mixing matrix can be as blockwise as possible.
We will briefly explain these concepts here.

{\bf $C$-parities.}
The $C$-parity operation over a length-$L$ Yang-Mills operators is to
reverse the color order of trace basis and  multiply a factor $(-1)^L$,
and the $C$-even (odd) operators are eigenstates of $C$-parity operation
with eigenvalues $+1$ ($-1$).
In  Yang-Mills theory $C$-parity is an exact symmetry, so
operators with different $C$-parities do not mix with each other
to all loop orders.

{\bf Helicity sectors.}
For a non-vanishing form factor at $d=4$ 
the helicity configurations of external particles can be classified into 
different types according to the numbers of $+$ and $-$ helicities they contain.
We construct the bases of physical operators according to the  helicity structures of their 
form factors. 
That is to say, the minimal form factor of each basis operator
does not vanish under only one pair of specific helicity configurations and its conjugate one,
and we say this operator belongs to such helicity sector. 

For example, the minimal form factor 
of the length-2 operator $\mathrm{tr}F^2$ survives under 
$(\pm,\pm)$ and vanishes under $(+,-)$, so we say $\mathrm{tr}F^2$ belongs to
helicity sector $(-,-)$. Here we select one representative
of the two conjugate configurations to label each helicity sector.

{\bf $D$-type operators.} 
Operators with total derivatives are a special class of operators which are usually called descendents. We will include them in our operator basis but take special care of them as $D$-type operators.
An operator is said to be of $D$-$(i,\alpha)$ type if it is an $i$th total derivative of a rank-$\alpha$ tensor operator. %
For example, at $\Delta_0=10$, the length-4 single-trace $C$-even evanescent sector is composed of three operators
\begin{align}
\label{eq:egforDtype}
&\mathcal{O}^e_{10;\mathrm{s}+;1}=
\partial_{\mu}\partial_{\nu} \frac{1}{16} \delta^{1234 \mu}_{5678 \nu}
\Big(4 \text{tr}( F_{12} F_{34}F_{78} F_{56} )
+2 \mathrm{tr}(F_{12} F_{56} F_{34} F_{78})\Big)\,,
\nonumber\\
&\mathcal{O}^e_{10;\mathrm{s}+;2}=
\partial_{\mu}\partial_{\nu} \frac{1}{16} \delta^{1234 \mu}_{5678 \nu}
\Big(4 \text{tr}(F_{12} F_{34}F_{78} F_{56}  )
-4 \mathrm{tr}(F_{12} F_{56} F_{34}F_{78} )\Big)\,,
\nonumber\\
& \mathcal{O}^e_{10;\mathrm{s}+;3}
 =   \frac{1}{2}
\delta^{1234\mu}_{5678 \nu}
\Big( \mathrm{tr}( D_{\nu}F_{12} F_{78}F_{34} D_{\mu}F_{56} )
-\frac{1}{2}  \mathrm{tr}( D_{\nu}F_{12}  F_{34} D_{\mu}F_{56}  F_{78} )
-\frac{1}{2} \mathrm{tr}(D_{\nu}F_{12} F_{78}  D_{\mu}F_{56} F_{34})
\Big)
\,.
\end{align}
By definition, the first two operators are of  $D$-(2,2) and the last one is of $D$-(0,0).
As discussed in \cite{Jin:2022qjc}, one can impose an ordering on the $D$-types as
\begin{align}
	&D\text{-}(i,\alpha)>D\text{-}(i',\alpha'),\ \text{if $i>i'$}\,,\\ &D\text{-}(i,\alpha)>D\text{-}(i,\alpha'),\ \text{if $\alpha<\alpha'$}\,.
\end{align}

A good property of this arrangement is that the renormalization matrix would be block upper triangular such that higher $D$-type operators do not mix with lower $D$-type ones. Therefore, to calculate anomalous dimensions, we only need to calculate the diagonal blocks, namely the sub-block of each $D$-type. And we can classify the anomalous dimensions according to $D$-types.

\section{ Gram matrices}
\label{sec:GramMatrix}

In this section, we consider the calculation of the Gram matrix associated with the two-point functions, 
which encodes  the information of the norms of operators. We show there are negative-norm states, indicating
the violation of unitarity of YM theory in non-integer spacetime dimensions.
In Section~\ref{sec:Gcomputation}, we introduce an efficient method for computing the Gram matrix using form factors.
Section~\ref{sec:Gresult} presents explicit results of Gram matrices for specific operator sectors,
which, in particular, demonstrate the existence of negative-norm states starting from dimension-12 operators.

\subsection{Calculation of Gram matrix} 
\label{sec:Gcomputation}

This section details the calculation of the Gram matrix.
We will consider the leading order of perturbation, where no interaction vertices are inserted.
Essentially, we deactivate the coupling constant and work with a ``free theory''.
The Gram matrix is denoted as $G_{ij}$ in the two-point Green function
\begin{align}
\label{eq:def}
\langle \symbolOa^{\dag}_i(x)\symbolOa_j(0)\rangle=\frac{G_{ij}}{|x^2|^{\Delta_{\symbolOa_i}}}\,.
\end{align}
The nonzero $G_{ij}$ requires operator $\symbolOa_i$ and $\symbolOa_j$ have the same canonical dimension and length.
This will be enough for the purpose of this paper to detect the negative-norm states, although the strategy can be generalized to higher-order correction.

Instead of the conventional computation through Wick contraction,
we propose a more efficient method based on form factors.
Briefly, we first work in momentum space by 
sewing two form factors to obtain a cut two-point Green function.
We then apply a Fourier transform to coordinate space by
translating the momenta back to derivatives,
multiplying the denominators of cut propagators back,
and finally performing differentiation and Lorentz contraction.

Below we start with a scalar theory in Section~\ref{sec:scalar} 
as a simpler example, which effectively demonstrates how to obtain the two-point functions
from the method based on unitarity cut.
Subsequently, we focus on our main interest, the Yang-Mills theory in Section~\ref{sec:YM}, 
where we introduce the more complex calculation steps due to the appearance of
particle polarization vectors and color factors.

\subsubsection{Warm up: scalar theory}
\label{sec:scalar}

Before the discussion on Yang-Mills operators, let us consider a simpler
sector composed of single-flavor scalar fields.

The steps of calculation are summarized as follows.

\begin{figure}[t!]
  \centering
  \includegraphics[scale=0.7]{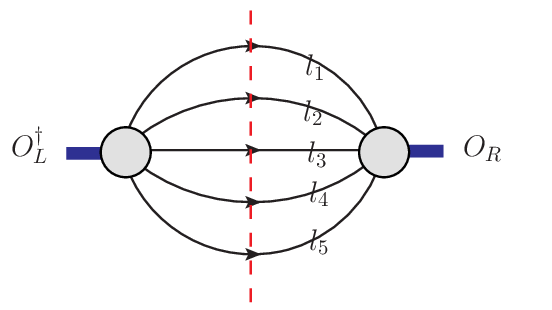}
  \caption{\label{fig:cutGreen} Cutting propagators makes the Green function factorize into a product of tree-level form factors.}
\end{figure}

We first consider the two-point Green function of length-$n$ operator $\symbolOa_L$ and $\symbolOa_R$
in momentum space. We impose unitarity cut $\frac{\mathbbm{i}}{l_i^2}\rightarrow \delta(l_i^2)$
as depicted in Figure \ref{fig:cutGreen}.
Resultantly this two-point Green function  factorizes
into a product of minimal form factors of  $\symbolOa_L$ and $\symbolOa_R$:%
\footnote{Here the cut of a two-point function in momentum space can be represented by the product of form factors by the optical theorem, see \emph{e.g.}~\cite{Nandan:2014oga}. Note also that the complex conjugate of the form factor is encoded in replacing ${\cal O}$ by its conjugate ${\cal O}^{\dag}$ and $p_i$ by $-p_i$.}
 \begin{align}
\label{eq:scalarpro0}
\langle\mathcal{O}_L^{\dag}\mathcal{O}_R\rangle^{(0)} \Big|_{\mathrm{cut}}=
\frac{1}{n!}\big(\mathcal{F}^{(0)}_{n,\mathcal{O}_L}(\{p_i\})\big)^*\times
\mathcal{F}^{(0)}_{n,\mathcal{O}_R}(\{p_i\})\,.
\end{align}
Here we label the length of $\symbolOa_L$ and $\symbolOa_R$ as $n$.
The cut two-point function only depends on kinematic variables $p_i\cdot p_j=s_{ij}/2$.

As a concrete example, let us consider the following two operators:
\begin{align}
\label{eq:slen4Op}
\symbolOa_{s1}:&=
 (\partial^\mu \partial^\nu\phi)^2
 ( \partial^\rho \partial^\sigma\phi)^2\,,
\nonumber\\
\symbolOa_{s2}:&=
 (\partial^\mu \partial^\nu\phi)
(\partial^\nu \partial^\rho\phi)( \partial^\rho \partial^\sigma\phi)
(\partial^\sigma\partial^\mu\phi)\,.
\end{align}
The form factors read from them are
\begin{align}
\label{eq:eg1FF}
\mathcal{F}^{(0)}_{4,\symbolOa_{s1}}&=\frac{1}{2}
s_{12}^2s_{34}^2+(\mbox{cyclic of } 2,3,4)\,,
\nonumber\\
\mathcal{F}^{(0)}_{4,\symbolOa_{s2}}&=\frac{1}{2}
s_{12}s_{23}s_{34}s_{14}+(\mbox{cyclic of } 2,3,4).
\end{align}
Due to bosonic symmetry, they are invariant under the permutation of $\{p_i\}$.
Plugging (\ref{eq:eg1FF}) into (\ref{eq:scalarpro0}), one gets the cut two-point function
\begin{align}
\label{eq:scalarpro1}
&\langle \symbolOa_{s1}^{\dag}\symbolOa_{s2}\rangle^{(0)} \Big|_{\mathrm{cut}}=
\frac{1}{4!} \mathcal{F}^{(0)}_{4,\symbolOa_{s1}^\dag }(\{-p_i\}) 
\times\mathcal{F}^{(0)}_{4,\symbolOa_{s2}}(\{p_i\})
\nonumber\\
&=\frac{1}{4!}
\Big(
\frac{1}{2}
s_{12}^2s_{34}^2+(\mbox{cyclic of } 2,3,4)
\Big)
\Big(\frac{1}{2}
s_{12}s_{23}s_{34}s_{14}+(\mbox{cyclic of } 2,3,4)
\Big)\,.
\end{align}

Next, we multiply the denominators of cut propagators back and
take Fourier transform back to coordinate space, \emph{i.e.}
translating each $p_i^\mu$ in the cut two-point function  to a  differential operator 
$-\mathbbm{i}\partial^{\mu}$
acting over the $i$-th scalar propagator $\langle \phi \phi \rangle_i$ where the label $i$
is temporarily introduced to distinguish the constituent scalar fields.
We place $\symbolOa_L$ at coordinates $x^\mu$ and $\symbolOa_R$ at the origin point, so
for general spacetime dimension $d$ the free propagator  equals $(-x^2+\mathbbm{i} 0)^{1-d/2}$.

Without loss of generality, let us take the first monomial in the expansion of (\ref{eq:scalarpro1}) as an example:
\begin{align}
\label{egs1}
&(p_1\cdot p_2)^3 (p_2\cdot p_3)( p_3 \cdot p_4)^3 (p_1\cdot p_4)
\nonumber\\
=\,&(p_1^{\mu_1}p_2^{\mu_1}p_1^{\mu_2}p_2^{\mu_2}p_1^{\mu_3}p_2^{\mu_3})
(p_2^{\mu_4}p_3^{\mu_4})
(p_3^{\mu_5}p_4^{\mu_5}p_3^{\mu_6}p_4^{\mu_6}p_3^{\mu_7}p_4^{\mu_7})
(p_1^{\mu_8}p_4^{\mu_8})
\nonumber\\
=\,&(p_1^{\mu_1} p_1^{\mu_2} p_1^{\mu_3}p_1^{\mu_8} )
(p_2^{\mu_1} p_2^{\mu_2} p_2^{\mu_3}p_2^{\mu_4})
(p_3^{\mu_5} p_3^{\mu_6} p_3^{\mu_7} p_3^{\mu_4})
( p_4^{\mu_5} p_4^{\mu_6} p_4^{\mu_7} p_4^{\mu_8}).
\end{align}
Note that in flat spacetime there is no need to distinguish the upper and lower Lorentz indices.

Translate momenta to differential operators according to the following rule:
\begin{align}
p_i^\mu \rightarrow -\mathbbm{i}\partial^\mu \mbox{ acting over }
\langle\phi \phi \rangle_i,\quad i=1,2,3,4\,.
\end{align}
So (\ref{egs1}) becomes a product of four symmetric tensors:
\begin{align}
\label{egs2}
\partial_{\mu_1\mu_2\mu_3\mu_8}\langle\phi \phi \rangle_1\times
\partial_{\mu_1\mu_2\mu_3\mu_4}\langle\phi \phi \rangle_2\times
\partial_{\mu_5\mu_6\mu_7\mu_4}\langle\phi \phi \rangle_3\times
\partial_{\mu_5\mu_6\mu_7\mu_8}\langle\phi \phi \rangle_4\,.
\end{align}
Other monomials in (\ref{eq:scalarpro1}) can be translated similarly.

To evaluate (\ref{egs2}) one performs differentiation on the propagators and then applies Lorentz contractions. The result is
\begin{align}
(d-2)^4(d-1)d^4(d+1)^2(d+2)^2(d^3+d^2-d+8)(-x^2)^{-4-2d}\,.
\end{align}

Combining the contributions from other monomials  in (\ref{eq:scalarpro1}), one has
\begin{align}
\langle \symbolOa_{s1}^{\dag}\symbolOa_{s2}\rangle&=
f(d) (-x^2)^{-4-2d}\,,
\nonumber\\
f(d)&=8(d-2)^4 (d-1)d^4(d+1)(d+2)(3d^5+8d^4+13d^2+54d+48)\,.
\end{align}
By definition (\ref{eq:def}), the Gram matrix element  $G_{\symbolOa_{s1} \symbolOa_{s2}}$ is equal to $f(d)$.

For scalar operators, one can introduce the concept of ``$\phi$-type'' as the list of
the number of derivatives for each constituent $\phi$ \cite{Hogervorst:2015akt}.
It is easy to see two operators (\ref{eq:slen4Op})
form the basis of $\phi$-type: $\{2,2,2,2\}$.
After calculating different operator contents, one obtains a $2\times 2$ matrix
\begin{align}
\label{eq:Gslen4}
G_{\{2,2,2,2\}}&=
\begin{pmatrix}
\langle \symbolOa^\dag_{s1}\symbolOa_{s1}\rangle & 
\langle \symbolOa^\dag_{s1} \symbolOa_{s2}\rangle \\
\langle \symbolOa^\dag_{s1} \symbolOa_{s2}\rangle & 
\langle \symbolOa^\dag_{s2} \symbolOa_{s2}\rangle
\end{pmatrix}
\\
&=8(d-2)^4 (d-1)d^4(d+1)(d+2)
\nonumber\\
&\times
\left(
	\begin{array}{c c}
	d(3d^4+14d^3+37d^2+58d+8)&
   3d^5+8d^4+13d^2+54d+48  \\
 3d^5+8d^4+13d^2+54d+48   &
    3 d^5+2 d^4-13d^3+16d^2+40d+48
	\end{array}
	\right).\nonumber
\end{align}

There exists a special combination $\symbolOa'_{s1}= 3 \symbolOa_{s1}-6 \symbolOa_{s2}$ that vanishes when $d= 2,3$, 
see also \cite{Hogervorst:2015akt}.
This operator is expected to be a  null state when $d= 2,3$, \emph{i.e.}
it is orthogonal to all the operators.
To see this, one can make a basis change
\begin{align}
\label{eq:trans1}
\{\symbolOa_{s1},\symbolOa_{s2}\} \rightarrow
\{3 \symbolOa_{s1}-6 \symbolOa_{s2},\symbolOa_{s2}\}\,.
\end{align}
Correspondingly, the Gram matrix should go through a congruent transformation
\begin{align}
\symbolOa_i\rightarrow \Xi_{ij}\symbolOa_j\,,\quad
G_{ij}\rightarrow \Xi_{ik}\Xi_{jl}G_{kl}\,,\quad
\Xi=\begin{pmatrix}3 & -6\\ 0 & 1\end{pmatrix}\,.
\end{align}
And (\ref{eq:Gslen4}) is transformed to
\begin{align}
\label{eq:Gslen4b}
&G'_{\{2,2,2,2\}}=8(d-2)^4 (d-1)d^4(d+1)(d+2)
\nonumber\\
&\times
\left(
	\begin{array}{c c}
	9(d-3)(d-2)(d-1)d(8+3d)&
   -3(d-3)(d-2)(8+3d)(1+d+d^2)  \\
  -3(d-3)(d-2)(8+3d)(1+d+d^2)  &
   3 d^5+2d^4-13d^3+16d^2-40d+48
	\end{array}
	\right).
\end{align}
It is clear that the first row and the first column of (\ref{eq:Gslen4b}) vanish at $d=2,3$.

We would like to mention that there is an alternative way without Fourier
transforming back to coordinate space, namely, one can evaluate the 
Feynman integral of the watermelon-like graph in Figure~\ref{fig:cutGreen} directly in momentum space
through integral reduction method like IBP \cite{Chetyrkin:1981qh,Tkachov:1981wb}. 
We make the parallel computation for several examples and find the results of the two calculations coincide.

\subsubsection{Yang-Mills theory}
\label{sec:YM}

For Yang-Mills operators with general form (\ref{eq:operator}), the calculation of
their Gram matrices are similar to those in scalar theory,   augmented by
additional steps stemming from the existence of color factors and polarization vectors.

The first step, similar to (\ref{eq:scalarpro0}), is to write down the cut two-point function 
$\langle\symbolOa^\dag_L \symbolOa_R\rangle|_{\mathrm{cut}}$
which is a product of the form factors of the operator $\symbolOa_L$ and $\symbolOa_R$, respectively.
At the leading order of perturbation, both of the two cut components are tree-level minimal form factors,
and left and right operators are required to have the same length. 
Taking single-trace operators as an example, one has%
\footnote{Similar to the scalar theory, taking the complex conjugate of the form factor of a Yang-Mills operator amounts to replacing ${\cal O}$ by ${\cal O}^{\dag}$, changing the sign of momenta and helicities.}  
\begin{align}
\label{eq:YMpro1}
\langle \symbolOa^\dag_L  \symbolOa_R\rangle^{(0)}\Big|_{\mathrm{cut}}&
=\sum_{\mathrm{helicty}}
\big(\mathbf{F}^{(0)}_{n;\symbolOa_L} (\{p_i\})\big)^*
\mathbf{F}^{(0)}_{n;\symbolOa_R} (\{p_i\})
\\
&=\sum_{\mathrm{helicity}}
\Big(\sum_{\sigma\in S_{n-1}}
\mathrm{tr}(1\sigma(2)\cdots\sigma(n) )
\mathcal{F}^{(0)}_{n;\symbolOa_L^{\dag}} (-p_1,-p_{\sigma(2)},\cdots,-p_{\sigma(n)})
\Big)
\nonumber\\
&\times\Big(\sum_{\tau\in S_{n-1}}
\mathrm{tr}(1\tau(2)\cdots\tau(n))
\mathcal{F}^{(0)}_{n;\symbolOa_R} (p_1,p_{\tau(2)},\cdots,p_{\tau(n)})
\Big)\,,\nonumber
\end{align}
where $\mathbf{F}$ and $\mathcal{F}$ denote  full color form factors
and color-ordered form factors respectively.

Compared to the previous scalar theory, a new complication in YM theory is to perform the helicity sum for the polarization vectors of the cut gluon states.
In general $d$ dimensions,  the helicity sum is evaluated through 
\begin{align}
\label{eq:polarsum1}
\sum_{\mathrm{helicity}}
e_i^{*,\mu}   e_i^\nu= \eta_{\mu\nu}-\frac{p_i^\mu q_i^\nu+p_i^\nu q_i^\mu}{p_i\cdot q_i}\,,
\end{align}
where $q_i$ are light-like reference momenta.
Since the operators are gauge invariant, the cut two-point functions
are independent of the choice of $\{q_i\}$.
As a comparison, when $d=4$, the color-ordered form factors $\mathcal{F}^{(0)}_{n;\symbolOa_L}$
and $\mathcal{F}^{(0)}_{n;\symbolOa_R}$ can be written in the form of
spinor helicity representation, and the polarization sum requires the left and right form factors
have the conjugate helicity configurations, \emph{e.g.}
\begin{align}
\label{eq:polarsum2}
\sum_{\{h_i\}} \mathcal{F}^{(0)}_{n;\symbolOa_L^\dag} (-p_n^{-h_n},\cdots,-p_2^{-h_2},-p_1^{-h_1})
\mathcal{F}^{(0)}_{2;\symbolOa_R} (p_1^{h_1},p_2^{h_2},\cdots,p_n^{h_n})\,.
\end{align}

In addition to the helicity sum, the sum of color degrees of freedom requires the contraction of color indices, which can be performed through the completeness relation of $\mathfrak{su}(N_c)$ Lie algebra
\begin{align}
\sum_a (T^a)^i_{\ j} (T^a)^k_{\ l}=\frac{1}{2}\delta^i_{l}\delta^j_{k}
-\frac{1}{2N_c}\delta^i_{j}\delta^k_{l}\,.
\end{align}
The highest $N_c$-order of $G_{ij}$ is the length of the operators, which
is easy to count from color graph analysis.

As the simplest example,
consider $\symbolOa_L=\symbolOa_R=\mathrm{tr}(F_{\mu\nu}F^{\mu\nu})=\mathrm{tr}(F^2)$. The form factor is
\begin{align}
\mathbf{F}^{(0)}_{2;\mathrm{tr}(F^2)}=\mathrm{tr}(T^{a_1}T^{a_2})
 \big[2(e_1\cdot e_2 )(p_1\cdot p_2)-2(e_1\cdot p_2)( e_2 \cdot p_1)\big]\,.
\end{align}
The result of color  contraction is
\begin{align}
\mathrm{tr}(T^{a_2}T^{a_1})\mathrm{tr}(T^{a_1}T^{a_2})=N_c^2-1\,.
\end{align}
After color contraction and polarization sum, the product (\ref{eq:YMpro1}) becomes
\begin{align}
\label{egYM1}
\langle\mathrm{tr}(F^2)\mathrm{tr}(F^2)\rangle^{(0)}\Big|_{\mathrm{cut}}&=\sum_{\mathrm{polarizations}}
2(N_c^2-1) \big[2(e_1\cdot e_2 )(p_1\cdot p_2)-2(e_1\cdot p_2)( e_2 \cdot p_1)\big]^2
\nonumber\\
&=8(N_c^2-1) (d-2)(p_1\cdot p_2)^2
=8(N_c^2-1) (d-2) p_1^\mu p_2^\mu p_1^\nu p_2^\nu\,.
\end{align}

The remaining steps are similar to the scalar theory case. To compute the Gram matrix element, one can
multiply the denominators of cut propagators back and 
take Fourier transform back to coordinate space.
For the above example, the final result of the Gram matrix element is
\begin{align}
\label{eq:normofF2}
G_{\mathrm{tr}(F^2)\mathrm{tr}(F^2)}=
 8 (d-2)^3(d-1)d(N_c^2-1) \,.
\end{align}

\subsection{Results of Gram matrices}
\label{sec:Gresult}

In this section, we present the result of Gram matrices for YM operators of
different canonical dimensions or lengths.
Section~\ref{sec:positive} collects the results of relatively low dimensional
operators where Gram matrices are all positive definite.
We use these examples to introduce the general structure of Gram matrices.
In Section~\ref{sec:negative} we represent
the Gram matrices of operators with dimension$\geq12$, where we see the appearance of negative norms at $4<d<5$.

We will mainly focus on the result of leading order in $N_c$ contributed
from single-trace operators, which is sufficient to illustrate the main physical features.
The results of full $N_c$ dependence are given in Appendix~\ref{ap:otherG} as well as in the ancillary file.

\subsubsection{Gram matrices with $\Delta_0 \leq 10$}
\label{sec:positive}

For the YM operators with mass dimension $\Delta_0<12$,%
\footnote{Note that we consider YM operators that are Lorentz scalar. If tensor operators are also allowed, the evanescent operators
as well as negative-norm states would appear at lower mass dimensions than 12.}
the Gram matrices considered are all positive definite at $4<d<5$.
These examples, though not violating unitarity, 
reveal some general properties of the Gram matrices, such as the block-diagonal structure for the operator basis respecting the classification of length, $C$-parity
and helicity structure.

\subsubsection*{Dimension 4}

The only operator with canonical dimension four is $\mathrm{tr}F^2$, and its norm has
been given in (\ref{eq:normofF2}) as
\begin{align}
G _{\mathrm{tr}(F^2),\mathrm{tr}(F^2)}
=8\, (N_c^2-1)  \,d(d-1)(d-2)^3  \,.
\end{align}

As explained in our previous work \cite{Jin:2020pwh},
at each given canonical dimension $\Delta_0$ there is only one length-2 operator, which
can be chosen as
the total derivative $(\partial^2)^{\kappa}\mathrm{tr}(F^2)$ for $\kappa=\frac{\Delta_0-4}{2}$,
and its norm is
\begin{align}
2^{6\kappa}
\frac{\Gamma(\frac{d}{2}+\kappa)\Gamma(\frac{d+1}{2}+\kappa )
\Gamma(\frac{d}{4}+\kappa+1)\Gamma(\frac{d}{4}+\kappa+\frac{1}{2})
}{\Gamma(\frac{d}{2})\Gamma(\frac{d+1}{2})
\Gamma(\frac{d}{4}+1)\Gamma(\frac{d}{4}+\frac{1}{2})}
G^{(0)}_{\mathrm{tr}(F^2),\mathrm{tr}(F^2)}\,.
\end{align}
The $\Delta_0$-dependent factor is a trivial effect of descendants
coming from the tensor contraction 
$(\partial_{\mu_1\cdots \mu_{\Delta_0-2}}(x^2)^{1-d/2})^2$,
which can also be obtained from the bubble integral in momentum space,
as mentioned at the end of Section~\ref{sec:scalar}.
The norm of $(\partial^2)^{\kappa}\mathrm{tr}(F^2)$ is proportional to
the norm of $(\partial^2)^{\kappa+1}\phi^2$.

\subsubsection*{Dimension 6  }

At canonical dimension 6 there is only one length-3 operator $\mathrm{tr}(F^3):=\mathrm{tr}(F_{12}F_{23}F_{13})$,
and its norm  is
\begin{align}
G _{\mathrm{tr}(F^3),\mathrm{tr}(F^3)}
=6\, (N_c^3-N_c)  \,d(d-1)(d-2)^4(3d-8)  \,.
\end{align}

\subsubsection*{Dimension 8 }

At dimension 8,
there are two length-3 operators
\begin{align}
\label{eq:dim8L3oper}
&
\mathcal{O}_{8;3;\alpha}=\frac{1}{6}\partial^2 \mathrm{tr}
(F_{12}\,F_{13}\,F_{23})\,,\quad
\mathcal{O}_{8;3;\beta}=\mathrm{tr}
(D_1F_{23}\,D_4F_{23}\,F_{14})
-\frac{1}{6}\partial^2 \mathrm{tr}
(F_{12}\,F_{13}\,F_{23})\,,
\end{align}
and four length-4 single-trace operators
\begin{align}
\label{eq:dim8L4stroper}
&\mathcal{O}^s_{8;4;\alpha;1}=\frac{1}{2}
\mathrm{tr}(F_{12}\,F_{12}\,F_{34}\,F_{34})
+\frac{1}{4}\mathrm{tr}(F_{12}F_{34}F_{12}F_{34})
+\mathrm{tr}(F_{12}\,F_{34}\,F_{23}\,F_{14})
\,,
\nonumber\\
&\mathcal{O}^s_{8;4;\alpha;2}=\frac{1}{2}\mathrm{tr}(F_{12}\,F_{12}\,F_{34}\,F_{34})
+\frac{1}{4}\mathrm{tr}(F_{12}\,F_{34}\,F_{12}\,F_{34})
+\mathrm{tr}(F_{12}\,F_{23}\,F_{34}\,F_{14})
\,,
\nonumber\\
&\mathcal{O}^s_{8;4;\gamma;1}=\frac{1}{4}\mathrm{tr}(F_{12}\,F_{34}\,F_{12}\,F_{34})
+\mathrm{tr}(F_{12}\,F_{34}\,F_{23}\,F_{14})
\,,
\nonumber\\
&\mathcal{O}^s_{8;4;\gamma;2}=\frac{1}{2}\mathrm{tr}(F_{12}\,F_{12}\,F_{34}\,F_{34})
-\frac{1}{4}\mathrm{tr}(F_{12}\,F_{34}\,F_{12}\,F_{34})
+\mathrm{tr}(F_{12}\,F_{23}\,F_{34}\,F_{14})
\,.
\end{align}
They have been grouped into different helicity sectors, denoted by symbol $\alpha,\beta,\gamma$.
Recall that the helicity sector is defined when taking the limit $d\rightarrow4$.

As mentioned before, the operators with different lengths are orthogonal at the leading order
of perturbation, so the Gram matrices of length-3 and length-4 can be written individually.
The length-3 operators are ordered as $\{\mathcal{O}_{8;3;\alpha},\mathcal{O}_{8;3;\beta}\}$,
which belong to $(-)^3$ and $(-)^2(+)$ sectors respectively.
The corresponding Gram matrix reads
\begin{align}
\label{eq:G083s}
G _{8;3}=
2\, (N_c^3-N_c) \,(d-2)^4(d+2)(d^2-1)d^2\times
\left(
\begin{array}{c;{2pt/2pt}c}
(3d-8)(3d+2) & -(d-4)(3d+2)
\\[0.2em]
 \hdashline[2pt/2pt]\rule{0pt}{0.9\normalbaselineskip}
-(d-4)(3d+2) & (d-2)(d+8)
\end{array}
\right).
\end{align}
The length-4 operators are ordered as $\{\mathcal{O}^s_{8;4;\alpha;1},
\mathcal{O}^s_{8;4;\alpha;2},\mathcal{O}^s_{8;4;\gamma;1},
\mathcal{O}^s_{8;4;\gamma;2}\}$, where the first two are of $(-)^4$ sector
and the last two are of $(-)^2(+)^2$ sector.
The Gram matrix at the leading order of $N_c$ reads
\begin{align}
\label{eq:G084s}
&G _{8;4}=2\,N_c^4\, (d-2)^5(d-1)d\times
\\
&{\small
\left(
\begin{array}{cc;{2pt/2pt}cc}
d^3+34d^2-220d+336 & 2(13d^2-78d+120) & -(d-4)(d^2-26 d+60) & 6(d-4)(d-2)
\\[0.2em]
2(13d^2-78d+120) & 2(d^3+21d^2-142d+216) & 6(d-4)(d-2) & -2(d-4)(d^2-23d+54)
\\[0.2em]
 \hdashline[2pt/2pt]\rule{0pt}{0.9\normalbaselineskip}
-(d-4)(d^2-26d+60) & 6(d-4)(d-2) & d^3+18 d^2-124d+192 & -6(d-4)(d-2)
\\[0.2em]
6(d-4)(d-2) & -2(d-4)(d^2-23d+54) & -6(d-4)(d-2) & 2(d^3+21d^2-142d+216)
\end{array} \right).}\nonumber
\end{align}

The vertical and horizontal dashed lines in (\ref{eq:G083s}) and (\ref{eq:G084s}) divide the helicity sectors.
From the off-diagonal blocks of both two matrices,
one can see that the helicity-crossing matrix elements all have a factor $(d-4)$ and vanish at $d=4$.
One can understand this from the calculation of the Gram matrix: 
in the polarization sum at $d=4$,
one requires the left and right form factors to have conjugate helicity configurations, as shown in (\ref{eq:polarsum2});
and the tree-level minimal form factors are non-zero only for two operators belonging to the same helicity sectors.

\subsubsection*{Dimension 10 }

Operators of canonical dimension 10 introduce some new features.
First, as mentioned in Section~\ref{sec:setup},  (Lorentz-invariant)
evanescent operators also begin to appear at this mass dimension.
Vanishing at $d=4$, these operators are supposed to be null states
under the inner product defined through Gram matrices.
Moreover, $C$-odd operators begin to appear.
$C$-parity is an exact symmetry and therefore
operators with different $C$-parities are orthogonal to each other.

Take dim-10 length-4 single-trace operators as an example.
They can be divided into 18 $C$-even operators and 6 $C$-odd ones, and 
the explicit operator bases are given in Appendix~\ref{app:opdim10}.
The Gram matrix elements between operators with different $C$-parities
are always zero, so we can consider them separately.
Due to the large size of the matrix, we collect the result in Appendix~\ref{app:gramdim10}.
The Gram matrices of $C$-even and $C$-odd sectors at $d=4-2\epsilon$
 are given by (\ref{eq:G0104s+}) and (\ref{eq:G0104s-}).
In the limit of $\epsilon\rightarrow 0$,
one can also check that operators belonging to different helicity sectors are
orthogonal to each other.

An important new feature arises, due to the existence of evanescent operators, \emph{i.e.}
the rows and columns representing the evanescent operators vanish
at $d=4$.
In other words, evanescent operators are \emph{null states} at $d=4$.%
\footnote{For a real symmetric matrix $G$, the counting of its
linearly independent positive/negative/null states is equivalent
to counting its positive/negative/zero eigenvalues. This is because one can
always diagonalize $G$ with an orthonormal matrix, so the signature of $G$ is equal to the signature of
its eigenvalues.  }
For example,  the diagonal elements of the three $C$-even
evanescent operators (\ref{eq:egforDtype}) are
\begin{align}
\label{eq:GofD10L4Eva}
&G^+_{11}=c_0 \times 1536 (3d+2)(3d+4),\nonumber\\
&G^+_{22}=c_0\times 1920 (3d+2)(3d+4),
\nonumber\\
&G^+_{33}=c_0\times 128 (d+2)(3d^2+33d+28),
\end{align}
where 
\begin{equation}
c_0=N_c^4 (d-4)(d-3)(d-2)^5(d-1)d^2(d+1) \,.
\end{equation}
These norms are zero at $d=4$ and positive at $4<d<5$.
The Gram matrices of both $C$-even and $C$-odd operators are positive definite
at $4<d<5$.

Similarly, the evanescent operators in the dim-10 length-5 sector are also 
null states, as shown in the Gram matrix (\ref{eq:G0105s}).
The matrix corresponding to length-5 operators is also positive definite.
Therefore, no negative-norm state is created from evanescent operators with
mass dimension 10.

\subsubsection{Negative-norm states for $\Delta_0 \geq 12$}
\label{sec:negative}

Starting from $\Delta_0 = 12$, the Gram matrices show a genuine new feature that there are negative-norm states. As we will see, this is because starting from canonical dimension 12, evanescent operators with rank-6 Kronecker symbol appear (such operators are called $\delta$-6 type).
The rank-6 Kronecker symbol makes them vanish at both $d=4$ and $d=5$.
They exist in both length-4 and length-5 sectors.

\subsubsection*{Dimension 12}
\label{sec:dim12}

Firstly consider length-4 operators.
There are 107 dim-12 length-4 single-trace operators,
among which 25  are evanescent, and we list them
in Appendix~\ref{app:opdim12}.
Among 25 evanescent operators, 24 are of $\delta$-5 type, and
the remaining one is of $\delta$-6 type, which reads
 \begin{align}
\label{eq:opd12delta6}
\symbolOa_b
=\partial_{\mu}\partial_{\nu}\Big[\delta^{12456\mu}_{3789\rho\nu}\Big(\text{tr}(D_{1}F_{23}F_{45}D_{6}F_{78}F_{9\rho})+\text{trace reverse})\Big)\Big]
\,. 
\end{align}

The full $107\times 107$ Gram matrix is given in the ancillary file. 
It has 25 null states when $d=4$, in agreement with
the counting of evanescent operators.
The row and column elements which involve $\mathcal{O}_b$  have an
overall factor $(d-4)(d-5)$.
Especially, the diagonal element reads
\begin{align}
\label{eq:opd12delta6b}
G_{bb}=
1152 N_c^4  (3d+8)(d+2)(d+1)d^5(d-1)^2 (d-2)^5(d-3)(d-4)(d-5)\,,
\end{align}
which is negative when $4<d<5$.
This implies (from Sylvester's law of inertia \cite{ostrowski1959quantitative})
that the Gram matrix of  length-4 dim-12 operators is not positive definite.
signaling the violation of unitarity of the theory at $4<d<5$.

Next, we turn to the length-5 cases.
There are 151 dim-12 length-5 single-trace operators, among which 61 are evanescent.
Among 61 evanescent operators, 53 are of $\delta$-5 type,
and 8 are of $\delta$-6 type. The explicit expressions of the eight $\delta$-6 operators can be found in \eqref{eq:dim12lengh5delta6-1}-\eqref{eq:dim12lengh5delta6-2} in Appendix~\ref{app:opdim12}.

 The full $151\times 151$ Gram matrix is given in the ancillary file. 
It has 61 null states when $d=4$, in agreement with
the counting of evanescent operators.
Similar to the length-4 cases, the rows and columns involving
$\delta$-6 operators have an overall factor  $(d-4) (d-5) $.
The number of $\delta$-6 operators equals the number of negative-norm states at $4<d<5$.

So far we have found negative-norm states at the level of mass dimension 12 
among both length-4 and length-5 operators, characterized by their $\delta$-6 structures.
As a comparison, in scalar theory, the $\delta$-6 operators should have at least
length six \cite{Hogervorst:2015akt}, which has been explained in  Section~\ref{sec:setup}.
Yang-Mills operators permit lower lengths for $\delta$-6 structures because
some Lorentz indices are shared by field strengths $F_{\mu\nu}$.
This feature is manifested by the calculation steps introduced in Section~\ref{sec:YM}.  
For the length-4 and length-5 Yang-Mills $\delta$-6 operators,
the factor $(d-5)$ in the Gram matrix elements emerges exactly after the polarization sum.

\subsubsection*{Higher dimensions}
\label{sec:dim1416}

We briefly discuss the operators with dimensions $\Delta_0>12$. We have explicitly computed the Gram matrices for length-4 single-trace operators
with canonical dimensions 14 and 16.
In Table~\ref{tab:signature},
we summarize the numbers of (linearly independent)
positive and negative-norm states  at $4<d<5$,
counted from positive and negative eigenvalues of Gram matrices.
The numbers of negative-norm states always equal
the numbers of linearly independent $\delta$-6 operators, which are classified
and counted in our early work \cite{Jin:2022ivc}.

The $\delta$-6 operator $\mathcal{O}_b$ given in (\ref{eq:opd12delta6}) implies that there exist negative-norm states for operators with arbitrarily higher canonical dimensions.
One can construct arbitrarily high dimensional $\delta$-6 operators by
inserting pairs of identical $D_\mu$ into four sites of  $\mathcal{O}_b$.
In momentum space, these additional $D_\mu$ pairs result in extra factors purely composed of
four-particle Mandelstam variables.
Such factors do not affect the process of polarization sum, and they do not contribute
to any  $(d-4)$ or $(d-5)$ because four-particle Mandelstam variables cannot form a quantity that vanishes at $d = 4$ or $d=5$.
As a result, the norms of the constructed operators also have a factor linear in both $(d-4)$ and $(d-5)$
and thus are negative at $4<d<5$.

\begin{table}[!t]
\centering
\begin{tabular}{|c|c|c|c|c|c| }
\hline
$\Delta_0$ & $~8~$ & $ ~10~$ & $ 12$ & $ 14$ & $ 16$\\
\hline
\hline
$N_+^{\rm p}=N^{\rm p}$  & 4 & 20 & 82 & 232 & 550  \\
\hline
$N_+^{\rm e}$  & 0 & 4 & 24 & 88 & 246  \\
\hline
$N_-^{\rm e}$  & 0 & 0 & 1 & 4 & 13  \\
\hline
\end{tabular}
\caption{\label{tab:signature} Counting of states with positive and negative norms for the single-trace length-4 basis up to mass dimension $16$.
$N_+$ and $N_-$ are the numbers of positive- and negative-norm states, respectively.
Superscript p and e denote physical and evanescent. }
\end{table}

\section{Complex anomalous dimensions}
\label{sec:complexAD}

The negative-norm states of Gram matrices shown in previous sections are calculated in free theory where no interacting vertices are involved. In this section, we show that complex anomalous dimensions emerge in the presence of interactions, which is new evidence of unitarity violation.

After a brief review of the computational framework of anomalous dimensions in Section~\ref{admethod}, we present the anomalous dimension results in pure YM theory in Section~\ref{theadresults}, where our focus is on the sectors containing complex anomalous dimensions. As an extension of pure YM theory, in Section~\ref{matter} we show that complex anomalous dimensions also exist for Yang-Mills-scalar theory which can have an IR fixed point. We finally comment on the anomalous dimension at higher loops in Section~\ref{subsec:highloopAD}.

\subsection{Computational setup}
\label{admethod}

As mentioned in Section~\ref{sec:setup}, the canonical dimensions of the operators receive quantum corrections, \emph{i.e.}~anomalous dimensions $\gamma$, due to the renormalization. 
A renormalized operator can be written as
\begin{align}
	\mathcal{O}_i=Z_i^j \mathcal{O}_{j,\text{b}}\,,
	\label{reO}
\end{align}
where $Z$ is the renormalization matrix and $\mathcal{O}_{j,\text{b}}$ are bare operators. 
Using the $Z$ matrix, the dilatation matrix $\mathbb{D}$ can be calculated according to
\begin{align}
	\mathbb{D}\equiv -\mu\frac{\text{d}Z}{\text{d}\mu}Z^{-1}\,.
	\label{overallgamma}
\end{align}
The anomalous dimensions $\gamma$ are defined to be the eigenvalues of $\mathbb{D}$. 
The goal of this section is to obtain one-loop anomalous dimensions from the one-loop dilatation matrix, which is simply related to the one-loop renormalization matrix as 
\begin{align}
	\mathbb{D}^{(1)}=2\epsilon Z^{(1)}\,.\label{D1}
\end{align}

\begin{figure}[t]
	\centering
	\includegraphics[scale=0.8]{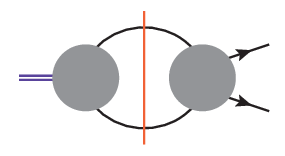}
	\caption{The double cut for one-loop minimal form factors.}
	\label{thedoublecut}
\end{figure}

We will follow the computational strategy using form factors which has been elaborated in \cite{Jin:2019opr, Jin:2020pwh, Jin:2022ivc, Jin:2022qjc}. Here we briefly review the main ideas. 
To calculate the one-loop renormalized operators, it is enough to calculate the minimal form factors, since there is no mixing between operators of different lengths. For a minimal form factor, the one-loop corrections can be expressed in terms of bubble integrals. In this case, the results can be computed efficiently using the double-cut shown in Figure~\ref{thedoublecut}. The cut form factor can be constructed by tree-level building blocks as
\begin{align}
	\mathcal{F}^{(1)}_b|_{\text{cut}}=\sum_{\text{polarization}} \mathcal{F}^{(0)} \times \mathcal{A}^{(0)}_4\,,
	\label{cutF}
\end{align}
where $\mathcal{F}^{(0)}$ is a tree-level minimal form factor and $\mathcal{A}^{(0)}_4$ is a tree-level four-point amplitude. 
We would like to emphasize that, since we aim to calculate the anomalous dimensions of the evanescent operators, we need to perform $d$-dimensional unitarity cuts, and the helicity sum for the polarization vectors of the internal cut gluons in $d$ dimensions is given by (\ref{eq:polarsum1}). The bare form form factor results are thus obtained in the conventional dimensional regularization (CDR) scheme.

Given the one-loop bare form factors, it is standard to extract the renormalization constants by subtracting the universal IR divergences from the renormalized form factors, which is
\begin{align}
	\mathcal{F}_{i}^{(1)}= \mathcal{F}_{b, i}^{(1)} + (Z_i^{\,j})^{(1)} \mathcal{F}_{j}^{(0)} \,.
\end{align}
The IR divergence of a $n$-point one-loop renormalized form factor in pure YM reads
\begin{align}
	\mathcal{F}_n^{(1)}\big{|}_{\text{IR}}=I_n^{(1)}(\epsilon)\mathcal{F}_n^{(0)}\label{Fir1}\,.
\end{align}
For example, in the planar limit, $I_n^{(1)}$ reads \cite{Catani:1998bh} 
\begin{align}
	I_{n}^{(1)}(\epsilon) = - {e^{\gamma_E \epsilon} \over \Gamma(1-\epsilon)} \bigg( \frac{N_c}{\epsilon^2} + \frac{\beta_0}{2 \epsilon} \bigg) \sum_{i=1}^n (-{s_{i,i+1}} )^{-\epsilon} \,. \label{thei1}
\end{align}
Our computation will follow closely  \cite{Jin:2022ivc, Jin:2022qjc}, where more details can be found.

The beta function of the Yang-Mills theory reads
\begin{equation} \label{eq:betaRG}
	\mu {d \alpha_s \over d\mu} = - 2 \epsilon \alpha_s - 2\beta_0 {\alpha_s^2 \over 4\pi} + {\cal O}(\alpha^3) \,, 
\end{equation}
with $\beta_0=\frac{11N_c}{3}$. 
Since $\beta_0>0$,  the pure YM theory has a Wilson-Fisher (WF) conformal fixed point 
at $d>4$ (namely $\epsilon<0$), which reads
\begin{equation}
	\alpha_* = - { 4\pi \epsilon \over \beta_0} + {\cal O}(\epsilon^2) \,,
\end{equation}
which is a UV fixed point. The anomalous dimension we consider will be understood as evaluated at the WF conformal fixed point.

Later in Section~\ref{matter}, we will also discuss the effect of matter fields by coupling scalar matter fields to the Yang-Mills theory. In that case, when the number of scalar fields is large enough, the fixed point may become an IR one and exist at $d<4$.

It is also instructive to comment on the connection between the Gram matrix $G$ and the dilation matrix ${\cal D}$. The matrix product of the dilatation matrix and the Gram matrix should be a symmetric matrix, see \emph{e.g.}~\cite{Hogervorst:2015akt}. This fact that $\mathcal{D}G$ is symmetric provides a useful consistency check for our calculation.
We give a discussion on the connection explicitly in Appendix~\ref{app:GandZ}.

\subsection{One-loop complex anomalous dimensions}
\label{theadresults}
We have shown in Section~\ref{sec:positive} that evanescent operators would lead to negative-norm states at non-integer spacetime dimensions, implying non-unitarity. In this subsection, we would show another signal of the breaking of the unitarity---the complex anomalous dimensions in the evanescent sector. The calculation of the anomalous dimensions is already described in the last subsection. The complex anomalous dimensions exist in both the full-color result and the planar-limit result. In this work, we would focus on the planar-limit result.\footnote{For a big enough $N_c$, the full-color result has a similar feature as the one in the planar-limit. We would give a detailed discussion of the small $N_c$ effect in another work.} 
In the planar limit, the calculation can be simplified such that the minimal form factor and four-point amplitude in \eqref{cutF} are all color-ordered. 
Since the length of an evanescent operator is at least four, our discussion starts with the length-4 operators. 

\subsubsection*{Length-4 operators}
Complex anomalous dimensions can exist only if there are both negative-norm and positive-norm states when $d$ is near four (see \emph{e.g.}~\cite{Hogervorst:2015akt} and our discussion in Appendix~\ref{sec:signature}). 
From Section~\ref{sec:dim12} we see that this condition is equivalent to the existence of one or more evanescent operators of $\delta$-$6$ type. The dimension of such an operator is at least 12, thus there is no complex anomalous dimension at dimensions lower than 12.

At dimension 12, the evanescent operators can be classified as in Table~\ref{l4d12 counting}, and their explicit expressions are given in Appendix~\ref{app:opdim12}. 
The $\delta$-$6$ operator (\ref{eq:opd12delta6})
only appears once in the $C$-even $D$-$(2,2)$ sector, which leads to a negative-norm state as discussed in Section~\ref{sec:dim12}. 
As mentioned in Section~\ref{sec:setup}, the anomalous dimensions are the eigenvalues of the sub-block matrices of each $D$-type.
Therefore we only need to consider the anomalous dimensions of the $C$-even $D$-(2,2)  sector. 

\begin{table}[t]
	\centering
	\begin{tabular}{|ccccccc|cccc|}
		\hline
		\multicolumn{7}{|c|}{C-even}                                                                                                                                                                                            & \multicolumn{4}{c|}{C-odd}                                                                                       \\ \hline
		\multicolumn{1}{|c|}{$D$-(4,2)}  & \multicolumn{1}{c|}{$D$-(3,3)}  & \multicolumn{1}{c|}{$D$-(2,0)}  & \multicolumn{2}{c|}{$D$-(2,2)}                                    & \multicolumn{1}{c|}{$D$-(1,1)}  & $D$-(0,0)  & \multicolumn{1}{c|}{$D$-(3,1)}  & \multicolumn{1}{c|}{$D$-(3,3)}  & \multicolumn{1}{c|}{$D$-(2,2)}  & $D$-(1,1)  \\ \hline
		\multicolumn{1}{|c|}{$\delta$-5} & \multicolumn{1}{c|}{$\delta$-5} & \multicolumn{1}{c|}{$\delta$-5} & \multicolumn{1}{c|}{$\delta$-6} & \multicolumn{1}{c|}{$\delta$-5} & \multicolumn{1}{c|}{$\delta$-5} & $\delta$-5 & \multicolumn{1}{c|}{$\delta$-5} & \multicolumn{1}{c|}{$\delta$-5} & \multicolumn{1}{c|}{$\delta$-5} & $\delta$-5 \\ \hline
		\multicolumn{1}{|c|}{2}          & \multicolumn{1}{c|}{1}          & \multicolumn{1}{c|}{1}          & \multicolumn{1}{c|}{1}          & \multicolumn{1}{c|}{7}          & \multicolumn{1}{c|}{2}          & 2          & \multicolumn{1}{c|}{1}          & \multicolumn{1}{c|}{2}          & \multicolumn{1}{c|}{3}          & 3          \\ \hline
	\end{tabular}
	\caption{The classification of length-4 dimension-12 evanescent single-trace operators. The last row shows the number of evanescent operators in $\delta$-5 or $\delta$-6 type.}
	\label{l4d12 counting}
\end{table}

The operators in this sector read
\begin{align}
	&\partial_\nu \partial_\rho\Big[\delta^{12456\nu }_{3789\mu \rho }\Big(\text{tr}(D_{1}F_{23}F_{45}D_{6}F_{78}F_{9\mu })+\text{trace reverse})\Big)\Big],\label{delta6D22}
	\\
	&\partial_\nu \partial_\rho\Big[\delta^{1}_{4} \delta^{2356\nu }_{789\mu \rho }\Big(\text{tr}(D_{1}F_{23}F_{45}D_{6}F_{78}F_{9\mu })+\text{trace reverse})\Big)\Big],
	\\
	&\partial_\nu \partial_\rho\Big[\delta^{1}_{4} \delta^{2356\nu }_{789\mu \rho }\Big(\text{tr}(D_{1}F_{23}D_{4}F_{56}F_{78}F_{9\mu })+\text{trace reverse})\Big)\Big],
	\\
	&\partial_\nu \partial_\rho\Big[\delta^{1}_{4} \delta^{2357\nu }_{689\mu \rho }\Big(\text{tr}(D_{1}F_{23}D_{4}F_{56}F_{78}F_{9\mu })+\text{trace reverse})\Big)\Big],
	\\
	&\partial_\nu \partial_\rho\Big[\delta^{1}_{4} \delta^{2367\nu }_{589\mu \rho }\Big(\text{tr}(D_{1}F_{23}F_{45}D_{6}F_{78}F_{9\mu })+\text{trace reverse})\Big)\Big],\\
	&\partial_\nu \partial_\rho\Big[\delta^{1}_{4} \delta^{2378\nu }_{569\mu \rho }\Big(\text{tr}(D_{1}F_{23}D_{4}F_{56}F_{78}F_{9\mu })+\text{trace reverse})\Big)\Big],
	\\
	&\partial_\nu \partial_\rho\Big[\delta^{1}_{5} \delta^{2347\nu }_{689\mu \rho }\Big(\text{tr}(D_{1}F_{23}D_{4}F_{56}F_{78}F_{9\mu })+\text{trace reverse})\Big)\Big],
	\\
	&\partial_\nu \partial_\rho\Big[\delta^{2}_{4}\delta^{1567\nu }_{389\mu \rho } \Big(\text{tr}(D_{1}F_{23}F_{45}D_{6}F_{78}F_{9\mu })+\text{trace reverse})\Big)\Big]\,.
	\label{lastD22}
\end{align}
where the ``trace reverse" of an operator $\text{tr}(X_1X_2\cdots X_n)$ means $\text{tr}(X_n\cdots X_2X_1)$. 
The $C$-even $D$-(2,2) sub-block $Z$-matrix reads
\begin{align}
	\label{D22length4talbe}	
    \frac{N_c}{\epsilon}
	\left(
	\begin{array}{cccccccc}
		\frac{8}{3} & -1 & \frac{13}{24} & 0 & -\frac{7}{3} & 0 & -\frac{7}{3} & -\frac{14}{3} \\
		\frac{1}{4} & \frac{41}{12} & -1 & -\frac{5}{12} & \frac{1}{3} & \frac{5}{24} & \frac{7}{6} & \frac{8}{3} \\
		0 & 2 & 0 & -\frac{8}{3} & 0 & \frac{2}{3} & 0 & -\frac{8}{3} \\
		0 & \frac{2}{3} & -\frac{7}{6} & 2 & 0 & \frac{2}{3} & 0 & 0 \\
		-\frac{1}{24} & \frac{1}{24} & \frac{3}{16} & -\frac{1}{24} & \frac{11}{3} & -\frac{5}{16} & -\frac{1}{4} & -1 \\
		0 & -\frac{2}{3} & -\frac{1}{3} & 0 & 0 & \frac{16}{3} & 0 & \frac{8}{3} \\
		-\frac{1}{12} & -\frac{3}{4} & -\frac{9}{32} & \frac{1}{4} & -\frac{29}{12} & \frac{5}{24} & \frac{5}{12} & -\frac{13}{6} \\
		\frac{5}{12} & \frac{1}{6} & -\frac{13}{64} & \frac{5}{12} & -\frac{3}{8} & -\frac{1}{8} & -\frac{5}{24} & \frac{47}{12} \\
	\end{array}
	\right)\,.
\end{align}
One can get the dilatation matrix according to \eqref{D1}. 
The anomalous dimensions are given as the roots of the eigen equation
\begin{align}
	&x^8-\frac{257 x^7}{6}+\frac{27281 x^6}{36}-\frac{191654 x^5}{27}+\frac{3001838 x^4}{81}-\frac{24366124 x^3}{243}\nonumber\\
	&+\frac{21495296 x^2}{243}+\frac{101673536 x}{729}-\frac{175325696}{729}=0\,.
	\label{dim12eqs}
\end{align}
There are two complex anomalous dimensions:
\begin{align}
\label{eq:sdata1}
	 6.6283\pm1.3284 \mathbbm{i} \,,
\end{align}
which are complex conjugate to each other since the dilatation matrix is real.

We also calculate the anomalous dimensions of the dimension-14 and -16 operators. The results are lengthy and we give them in the ancillary file. 
The counting of anomalous dimensions is given in Table~\ref{length4 counting}, which is an extension of Table \ref{tab:signature}.
We see that the number of complex anomalous dimensions grows with classical dimensions. 
An interesting observation is  
that up to dimension 16, the numbers of complex anomalous dimensions are always twice of
the numbers of the $\delta-6$ operators (negative-norm states when $4<d<5$).

\begin{table}[!t]
\centering
\begin{tabular}{|c|c|c|c|c|c| }
\hline
$\Delta_0$ & $~8~$ & $ ~10~$ & $ 12$ & $ 14$ & $ 16$\\
\hline
\hline
$N_+^{\rm p}=N^{\rm p}$  & 4 & 20 & 82 & 232 & 550  \\
\hline
$N_+^{\rm e}$  & 0 & 4 & 24 & 88 & 246  \\
\hline
$N_-^{\rm e}$  & 0 & 0 & 1 & 4 & 13  \\
\hline
$N_{\gamma\textrm{-complex}}$  & 0 & 0 & $1\times2$ & $4\times2$ & $13\times2$  \\
\hline
\end{tabular}
\caption{\label{length4 counting} The counting of our results for length-4 dimension-12,14 and 16 length-4 single-trace operators, including the numbers of operators, negative-norm
states and complex anomalous dimensions.}
\end{table}

\subsubsection*{Length-5 operators}

We present the anomalous dimensions of length-5 dimension-12 operators as a higher-length example. An overall classification of the evanescent operators is given in Table~\ref{length5 counting}. Below we focus on the anomalous dimensions of the $C$-even $D$-$(2,2)$ sector.

The six operators in this sector read 
\begin{align}
	&\partial_\nu \partial_\rho\Big[\delta^{12345\nu }_{6789\mu \rho }\Big(\text{tr}(F_{12}F_{34}F_{56}F_{78}F_{9\mu })-\text{trace reverse})\Big)\Big],\label{l5o6a}
	\\
	&\partial_\nu \partial_\rho\Big[\delta^{12357\nu }_{4689\mu \rho }\Big(\text{tr}(F_{12}F_{34}F_{56}F_{78}F_{9\mu })-\text{trace reverse})\Big)\Big],\\
	&\partial_\nu \partial_\rho\Big[\delta^{1}_{3} \delta^{2456\nu }_{789\mu \rho }\Big(\text{tr}(F_{12}F_{34}F_{56}F_{78}F_{9\mu })-\text{trace reverse})\Big)\Big],
	\\
	&\partial_\nu \partial_\rho\Big[\delta^{1}_{3} \delta^{2457\nu }_{689\mu \rho }\Big(\text{tr}(F_{12}F_{34}F_{56}F_{78}F_{9\mu })-\text{trace reverse})\Big)\Big],\\
	&\partial_\nu \partial_\rho\Big[\delta^{1}_{3} \delta^{2567\nu }_{489\mu \rho }\Big(\text{tr}(F_{12}F_{34}F_{56}F_{78}F_{9\mu })-\text{trace reverse})\Big)\Big],\\
	&\partial_\nu \partial_\rho\Big[\delta^{1}_{5} \delta^{2346\nu }_{789\mu \rho }\Big(\text{tr}(F_{12}F_{34}F_{56}F_{78}F_{9\mu })-\text{trace reverse})\Big)\Big]\,.\label{l5o6}
\end{align}
The corresponding $Z$-matrix is
\begin{align}\label{eq:Z1dim12Eva}
\frac{N_c}{\epsilon}
	\left(
	\begin{array}{cccccc}
		-\frac{11}{2} & \frac{4}{3} & -\frac{13}{6} & 3 & \frac{7}{3} & 0 \\
		-\frac{13}{6} & \frac{23}{6} & -\frac{4}{3} & \frac{4}{3} & \frac{7}{3} & 0 \\
		\frac{4}{15} & \frac{2}{5} & -\frac{9}{4} & \frac{9}{2} & \frac{3}{2} & -\frac{3}{4} \\
		-\frac{4}{15} & \frac{3}{5} & \frac{3}{8} & \frac{35}{12} & -\frac{17}{12} & -\frac{7}{24} \\
		-\frac{2}{15} & \frac{2}{15} & -\frac{29}{12} & -\frac{1}{6} & 4 & -\frac{3}{4} \\
		\frac{22}{15} & -\frac{4}{5} & \frac{13}{6} & -\frac{19}{3} & -3 & -\frac{1}{3} \\
	\end{array}
	\right)\,.
\end{align}
The anomalous dimensions are the roots of the equation
\begin{align}
	x^6-\frac{16 x^5}{3}-\frac{1099 x^4}{9}+\frac{16240 x^3}{27}+\frac{78193 x^2}{27}-\frac{1511744 x}{243}-\frac{4935587}{243}=0\,.
\end{align}
There is one pair of complex roots:
\begin{align}
	 8.6383 \pm0.84161 \mathbbm{i}  \,.
\end{align}
A full calculation of the length-5 dimension-12 anomalous dimensions shows that there exist 7 pairs of complex anomalous dimensions, implying that the non-unitarity also exists for length-5 operators. In this case, the number of complex anomalous dimensions is not twice the number of the negative-norm states.

\begin{table}[t]
	\centering
	\begin{tabular}{|ccccccc|cccc|}
		\hline
		\multicolumn{7}{|c|}{$C$-even}                                                                                                                                                                   & \multicolumn{4}{c|}{$C$-odd}                                                                             \\ \hline
		\multicolumn{1}{|c|}{$D$-(2,0)} & \multicolumn{2}{c|}{$D$-(2,2)}                            & \multicolumn{2}{c|}{$D$-(1,1)}                            & \multicolumn{2}{c|}{$D$-(0,0)}       & \multicolumn{1}{c|}{$D$-(2,2)} & \multicolumn{2}{c|}{$D$-(1,1)}                            & $D$-(0,0) \\ \hline
		\multicolumn{1}{|c|}{$\delta$-5}    & \multicolumn{1}{c|}{$\delta$-6} & \multicolumn{1}{c|}{$\delta$-5} & \multicolumn{1}{c|}{$\delta$-6} & \multicolumn{1}{c|}{$\delta$-5} & \multicolumn{1}{c|}{$\delta$-6} & $\delta$-5 & \multicolumn{1}{c|}{$\delta$-5}    & \multicolumn{1}{c|}{$\delta$-6} & \multicolumn{1}{c|}{$\delta$-5} & $\delta$-5    \\ \hline
		\multicolumn{1}{|c|}{2}         & \multicolumn{1}{c|}{2}      & \multicolumn{1}{c|}{4}      & \multicolumn{1}{c|}{2}      & \multicolumn{1}{c|}{12}     & \multicolumn{1}{c|}{2}      & 12     & \multicolumn{1}{c|}{2}         & \multicolumn{1}{c|}{2}      & \multicolumn{1}{c|}{14}     & 7         \\ \hline
	\end{tabular}
	\caption{The counting of length-5 dimension-12 evanescent single-trace operators.}
	\label{length5 counting}
\end{table}

\subsection{Operators in the Yang-Mills-scalar theory}\label{matter}

In this subsection, we couple the Yang-Mills theory to a set of scalar fields and define the Lagrangian as
\footnote{To give a simple example, we do not add a $\phi^4$ interaction in the Lagrangian. In this case, the $\phi^4$ interaction would appear due to renormalization and has no effect at the one-loop order.}
\begin{align}
 {\cal L}_{\rm YMS} = -{1\over2} {\rm tr}(F_{\mu\nu}F^{\mu\nu})+\sum_{i=1}^{N_f}\frac{1}{2}D^\mu\phi^a_i D_\mu\phi^a_i \,,
\end{align}
where the superscript ``$a$" is the color index and the subscript ``$i$" the flavor index. 
The scalar fields are in adjoint representation and the action of the covariant derivative on a scalar field reads
\begin{align}
	D_\mu \phi^a_i=\partial_\mu \phi^a_i+g f^{abc}A_\mu^b\phi^c_i\,.
\end{align}
The theory will be referred to as the Yang-Mills-scalar (YMS) theory. 
The one-loop beta function:
\begin{align}
	\beta^{\text{YMS}}_0=N_c(\frac{11}{3}-\frac{N_f}{6})\,.
	\label{SQCDbeta}
\end{align} 
One can see that when $N_f>22$ the one-loop beta function changes its sign and when $N_f=22$ the one-loop beta function vanishes.
Below we show that the complex anomalous dimensions exist in this theory for general $N_f$, including large $N_f$ where the beta function $\beta_0$ becomes negative.

From the analysis of Kronecker symbols, one can derive that the evanescent operators including scalar fields begin to appear from classical dimension 12 and are at least length-4 operators. In the YMS, there are two types of length-4 evanescent operators at dimension 12:
\begin{align}
	&O^{\text{eva}}_{1,ij}=\partial_9 \partial_\mu\Big[\delta^{12459}_{3678\mu }\Big(\text{tr}(D_{1}\phi_i D_{23}\phi_j D_{4}F_{56}F_{78})\Big)\Big],\label{scalar1}
	\\
	&O^{\text{eva}}_{2,ij}=\partial_9 \partial_\mu\Big[\delta^{12359}_{4678\mu }\Big(\text{tr}(D_{1}\phi_i D_{2}F_{34}D_{56}\phi_j F_{78})\Big)\Big]\,,\label{scalar2}
\end{align}
where $i$ and $j$ are flavor indices. The new evanescent operators are in the $C$-even $D$-(2,2) sector, where there is a pair of complex anomalous dimensions in the pure YM theory as shown in Section~\ref{theadresults}. {One can further define two flavor-singlet scalar
\begin{align}
	O_{1,0}\equiv\frac{1}{N_f}\sum_{i}O_{1,ii},\qquad O_{2,0}\equiv\frac{1}{N_f}\sum_{i}O_{2,ii}\,.
\end{align}
These two operators and the gluonic operators in \eqref{delta6D22}-\eqref{lastD22} form the subset of flavor-singlet operators in the $D$-(2,2) sector. Since a flavor-singlet operator would not mix with a non-singlet one, it is enough to consider the flavor-singlet operators to study the effect of scalar fields on the complex anomalous dimensions.}

The calculation of anomalous dimensions is similar to the one in the YM theory as described in Section~\ref{admethod} except for two differences. The first one is the one-loop factor in the IR divergence:
\begin{align}
	I_{n}^{'(1)}(\epsilon) = -\frac{e^{\gamma_E\epsilon}}{\Gamma(1-\epsilon)}\sum_{i=1}^{n}\left(\frac{N_c}{\epsilon^2}+\frac{\gamma_i+\gamma_{i+1}}{2\epsilon}\right)(-s_{i,i+1})^{-\epsilon} \,, \label{thei1SQCD}
\end{align}
where $\gamma_i$ depends on the particle-type of the $i$th external leg: 
\begin{equation}
\gamma_g=N_c(\frac{11}{6}-\frac{N_f}{12}),\qquad \gamma_\phi=2N_c \,.
\end{equation}
The other one is the change of beta function as in \eqref{SQCDbeta}.

The $Z$-matrix of the flavor-singlet operators reads
\begin{align}\frac{N_c}{\epsilon}
\left(
\begin{array}{cccccccccc}
	\frac{8}{3} & -1 & \frac{13}{24} & 0 & -\frac{7}{3} & 0 & -\frac{7}{3} & -\frac{14}{3} & -8 N_f & 0 \\
	\frac{1}{4} & \frac{41}{12} & -1 & -\frac{5}{12} & \frac{1}{3} & \frac{5}{24} & \frac{7}{6} & \frac{8}{3} & 4 N_f & 0 \\
	0 & 2 & 0 & -\frac{8}{3} & 0 & \frac{2}{3} & 0 & -\frac{8}{3} & -\frac{16 N_f}{3} & 0 \\
	0 & \frac{2}{3} & -\frac{7}{6} & 2 & 0 & \frac{2}{3} & 0 & 0 & 0 & 0 \\
	-\frac{1}{24} & \frac{1}{24} & \frac{3}{16} & -\frac{1}{24} & \frac{11}{3} & -\frac{5}{16} & -\frac{1}{4} & -1 & -2 N_f & 0 \\
	0 & -\frac{2}{3} & -\frac{1}{3} & 0 & 0 & \frac{16}{3} & 0 & \frac{8}{3} & \frac{16 N_f}{3} & 0 \\
	-\frac{1}{12} & -\frac{3}{4} & -\frac{9}{32} & \frac{1}{4} & -\frac{29}{12} & \frac{5}{24} & \frac{5}{12} & -\frac{13}{6} & -\frac{10 N_f}{3} & 0 \\
	\frac{5}{12} & \frac{1}{6} & -\frac{13}{64} & \frac{5}{12} & -\frac{3}{8} & -\frac{1}{8} & -\frac{5}{24} & \frac{47}{12} & -\frac{5 N_f}{3} & 0 \\
	0 & 0 & \frac{5}{384} & 0 & -\frac{5}{48} & 0 & -\frac{5}{48} & -\frac{5}{24} & \frac{N_f}{6}+4 & -\frac{2}{3} \\
	0 & 0 & 0 & 0 & 0 & 0 & 0 & 0 & -\frac{1}{3} & \frac{N_f}{6}+\frac{11}{3} \\
\end{array}
\right)\,,
\end{align}
where the first eight rows (columns) correspond to the gluonic operators in \eqref{delta6D22}$\sim$\eqref{lastD22} and the last two rows (columns) correspond to $O_{1,0}$ and $O_{2,0}$.

The dilatation matrix is given by $\mathcal{D}^{(1)}=2\epsilon Z^{(1)}$. The anomalous dimensions are the roots of the eigenvalue equation
\begin{align}
	\sum^{10}_{i=0} g_i x^i=0\,,
	\label{thexequation}
\end{align}
where each $g_i$ is a polynomial of the parameter $N_f$. The number of real roots for a uni-variant polynomial equation can be calculated via Sturm's theorem \cite{sturm2009memoire}. (A detailed description of the analysis is given in the Appendix~\ref{root analysis}.) It turns out that for any positive integer $N_f$, there would always be one pair of complex anomalous dimensions in this sector. We provide some representative data in Table~\ref{scalar complex}.
\begin{table}[t]
	\centering
	\begin{tabular}{|c|c|c|c|c|}
		\hline
	$N_f$	& 1 & 10  & $10^3$ & $10^6$ \\ \hline
	Complex ADs	& 6.4078 $\pm $ 1.3041 $\mathbbm{i}$ & 6.1163 $\pm $ 1.3796 $\mathbbm{i}$ & 6.0214 $\pm $ 1.3941 $\mathbbm{i}$ & 6.0202 $\pm $ 1.3942 $\mathbbm{i}$ \\ \hline
	\end{tabular}
	\caption{The pairs of complex ADs for different $N_f$. }
	\label{scalar complex}
\end{table}

\subsection{On higher-loop corrections}\label{subsec:highloopAD}
So far we have considered the one-loop level and the complex ADs imply the unitarity violation. In this subsection, we give some discussion about higher-loop corrections.

The ADs are the eigenvalues of the dilatation matrices, \emph{i.e.} they are solutions of
\begin{align}
	\det{(\mathbb{D}-\gamma\mathbb{I})}=0\,,
	\label{eq:eigengeneral1}
\end{align}
where $\mathbb{I}$ denotes the identity matrix.
Both the dilatation matrix and anomalous dimension can be expanded with respect to the perturbative coupling $\alpha_s$ as
\begin{equation}
	\mathbb{D}=\sum_{j=1}^\infty \mathbb{D}^{(j)} \big( \frac{\alpha_s}{4\pi} \big)^j\,, \qquad
	\gamma=\sum_{j=1}^\infty \gamma^{(j)} \big(\frac{\alpha_s}{4\pi} \big)^j\,.	
\end{equation}
A useful fact is that the perturbative matrix elements of $\mathbb{D}^{(j)}$, which are related to the renormalization matrix $Z$ via \eqref{overallgamma},  are all real numbers in our choice of operator bases.
The one-loop ADs are solutions of the leading order equation
\begin{align}
	\det{(\mathbb{D}^{(1)}-\gamma^{(1)}\mathbb{I})}=0\,,
	\label{LOeigen}
\end{align}
which we have discussed in detail above in this section.

Let us suppose the dilatation matrix $\mathbb{D}$ is obtained up to $L$-loop order. 
For any specific AD $\gamma_n^{(j)}$, 
it must satisfy the eigenvalue equation
\begin{align}
	\det{\Big[\sum_{j=1}^{L} \big(\mathbb{D}^{(j)}- \gamma_n^{(j)}\mathbb{I} \big) \big( \frac{\alpha_s}{4\pi} \big)^{j-1}\Big]}=0\,.
	\label{eigengeneral}
\end{align}
Expanding \eqref{eigengeneral}, at $\mathcal{O}(\alpha_s^{L-1})$, one obtains the following equation linear in $\gamma^{(L)}_n$
\begin{align}
	\mathcal{C}_n \gamma^{(L)}_n=X_n^{(L)}\,,
	\label{orderk}
\end{align}
where the coefficient $\mathcal{C}_n$ can be given in terms of one-loop ADs as
\begin{align}
	\mathcal{C}_n=\prod_{m\neq n} (\gamma^{(1)}_n-\gamma^{(1)}_m)\,,
	\label{cos of next gamma}
\end{align}
and $X_n^{(L)}$ includes all terms of the LHS of Eq.\,\eqref{eigengeneral} without $\gamma_n^{(L)}$ at $\mathcal{O}(\alpha^{L-1})$.

Using \eqref{orderk} one can prove that, if the one-loop AD $\gamma_n^{(1)}$ is real and non-degenerate, $\gamma_n^{(L)}$ is real for any $L$.
Let us assume that $\gamma^{(j)}$ ($j<L$) are all real. Since $\mathbb{D}^{(1)}$ is real and $\gamma_n^{(1)}$ is real and non-degenerate, $\mathcal{C}_n$ in~\eqref{cos of next gamma} must be an non-vanishing real number. Since the LHS of \eqref{eigengeneral} is a determinant of a real matrix (up to the single variable $\gamma_n^{(L)}$), $X_n^{(L)}$ must be real too. Thus $\gamma_n^{(L)}$ is real according to \eqref{orderk}. 
Given that the starting point $\gamma_n^{(1)}$ is real, we conclude that $\gamma_n^{(L)}$ is real for any $L$ by mathematical induction.
As an example, since all one-loop planar ADs for dimension-10 operators are real and non-degenerate, we conclude that their ADs are real in any order.

For a complex one-loop $\gamma_n^{(1)}$, the above analysis is not sufficient to determine whether $\gamma_n^{(L)}$ is complex or real since both $\mathcal{C}_n$ and $X^{(L)}_n$ are complex in \eqref{orderk}. Despite this uncertainty, in a perturbative calculation, if $\gamma_n^{(1)}$ is complex, $\sum_{i=1}^{L}\gamma_n^{(i)}(\frac{\alpha_s}{4\pi})^i$ must be complex up to any order $L$. Therefore, a complex one-loop AD is sufficient to imply the unitarity violation.

\section{Summary and discussion}
\label{sec:discuss}

Evanescent operators, which are absent in four-dimensional spacetime but manifest in non-integer dimensions, play a crucial role in quantum field theory. 
Our investigation into gauge theories, including pure Yang-Mills (YM) theory and YM theory coupled with scalar fields, reveals that these operators cause negative-norm states and complex anomalous dimensions.
This pattern of unitarity violation, initially observed in scalar $\phi^4$ theory \cite{Hogervorst:2014rta, Hogervorst:2015akt} (see also the evidence of negative norms in fermionic-type theory \cite{Ji:2018yaf}), now appears to be a widespread phenomenon in quantum field theory at non-integer spacetime dimensions.

Given the necessity of calculations in full $d$ dimensions, rigorous cross-verification is essential.
We have validated our computations through several non-trivial checks, summarized as follows.
\begin{itemize}
\item 
First, the Gram matrices exhibit the correct zero matrix elements as predicted by operator properties,
such as the orthogonality between operators of different $C$-parities (at general $d$) 
and different helicity sectors (defined in the $d=4$ limit), 
as well as zero rows and columns between evanescent operators of different ranks of the Kronecker symbols.
\item 
Second, the one-loop form factors display correct IR divergences,
and the one-loop renormalization matrices demonstrate accurate mixing behaviors as predicted by the properties of operators,
including non-mixing from higher to lower $D$-type operators
and between operators with different  $C$-parities and helicity sectors.
\item 
Finally, the product of the Gram matrix and dilation matrix in a given sector shows the correct symmetry property,
as detailed in Appendix~\ref{app:GZ symmetric}.
\end{itemize}
We now explore the implications and potential generalizations of our findings.

First of all, the unitarity of gauge theories in four dimensions should remain intact despite the conclusions drawn from our study.
In a strict four-dimensional context, evanescent operators are nonexistent, and their involvement in dimensional regularization does not alter this fact.
From our perspective, one starts with the initial theory defined in $d=4-2\epsilon$ which is non-unitary.
Yet, as one approaches the limit $\epsilon\rightarrow0$, a unitary four-dimensional theory is expected to emerge. 
It is crucial that the spectral curves of physical and evanescent operators remain sufficiently distinct, allowing for the seamless removal of evanescent operators as one reaches the four-dimensional limit.
Our explicit calculations show that the eigenstates corresponding to the complex eigenvalues invariably have zero norm; thus, in the four-dimensional limit, such zero-norm states should be decoupled.
See also \cite{Hogervorst:2014rta, Hogervorst:2015akt} for discussion on this point.

In computing the one-loop anomalous dimensions, we have employed the unitarity-cut method \cite{Bern:1994cg,Bern:1994zx,Britto:2004nc}. 
Concerns may arise regarding the reliability of results derived from a method named `unitarity' in the context of a theory that exhibits unitarity violations.
However, it is important to clarify that the unitarity-cut method, as applied in modern computations, does not directly pertain to the imaginary part of the S-matrix.
Instead, it is a technique for determining the ``full integrand" of amplitudes or form factors, producing results that are equivalent with those obtained via traditional Feynman diagram calculations. 
This is particularly evident in the one-loop computations presented in this paper, where no ghost particles appear even in the Feynman-diagram approach.

Conformal Field Theories (CFTs) in non-integer dimensions may be studied using the bootstrap method \cite{Rattazzi:2008pe}. 
In particular, the conformal blocks can be defined in general dimensions \cite{Dolan:2000ut, Dolan:2003hv}.
However, a fundamental assumption of the standard bootstrap method is the unitarity of the underlying theory. 
Given the non-unitary nature of the QFTs in non-integer dimensions, as shown in this paper and also previous works \cite{Hogervorst:2015akt, Ji:2018yaf}, 
the applicability of the standard bootstrap method warrants scrutiny.
Interestingly, it was found in \cite{El-Showk:2013nia} that the bootstrap for the scalar theory has produced consistent results. 
This may be explained by the fact that in the scalar theory, unitarity violation occurs only at very high-dimensional states, so the unitarity-violating effect is well suppressed \cite{Hogervorst:2014rta, Hogervorst:2015akt}. See also other bootstrap studies in non-integer dimensions \cite{Codello:2014yfa, Golden:2014oqa, Chester:2014gqa}.
The work presented here reveals that unitarity-violating states arise at considerably lower dimensions in YM theories. 
Exploring their effect within the bootstrap paradigm would be an interesting avenue for future research.

Exploring gravitational theories presents another intriguing avenue for generalization. 
Analogous evanescent operators can be constructed in gravity by substituting gauge field strength tensors $F_{\mu\nu}$ with curvature $R_{\mu\nu\rho\sigma}$ or $R_{\mu\nu}$. 
It would be interesting to study the unitarity-violating effect in gravitational contexts. 
In particular, the holographic principle states that a gauge theory in $d$ dimensions would have a holographic dual of gravity (string) theory in $d+1$ dimensions \cite{tHooft:1993dmi, Susskind:1994vu, Maldacena:1997re}. It would be highly interesting to see in which sense the non-unitary gauge theory in non-integer $d$ dimensions implies that unitarity violation in the dual gravity theory. 
Since this is a weak/strong coupling duality, it may also reveal properties of gauge theory in the strong coupling regime.

Finally, the physical interpretation of non-integer dimensions in spacetime remains an open question.
One avenue for exploration is the use of a fractal lattice model to represent spacetime.
By constructing such a lattice and subsequently transitioning to a continuum limit, 
one could potentially arrive at a theory intrinsically defined within fractional dimensions.
This methodology echoes the approach taken in the analysis of critical phenomena on fractal lattices; see \emph{e.g}, \cite{Gefen:1980lho}.

\section*{Acknowledgements}
We would like to thank Bo Feng, Tao Shi, and Junbao Wu for discussions. 
This work is supported in part by the National Natural Science Foundation of China (Grants No.~11935013, 12175291, 12047503) and by the CAS under Grants No.~YSBR-101.
We also thank the support of the HPC Cluster of ITP-CAS.

\appendix

\section{Relations of the Gram matrix and the dilatation matrix}

In this appendix, we discuss some useful connections between the Gram matrix and the dilatation matrix.

\subsection{Anomalous dimensions from two-point functions}
\label{app:GandZ}

In Section~\ref{sec:complexAD}, we have defined the renormalization matrix $Z$ as
\begin{align}
	\mathcal{O}_i=Z_i^{~j} \mathcal{O}_{j,\text{b}}\,.
\end{align}
From the UV divergences of one-loop form factors, we extract the one-loop renormalization matrices, and so are the one-loop dilatation matrix 
\begin{align}
	\mathbb{D}^{(1)}=2\epsilon Z^{(1)}\,.
\end{align} 
The one-loop anomalous dimension is given by the eigenvalues of $\mathbb{D}^{(1)}$.

Alternatively, the one-loop anomalous dimensions can also be obtained from computing the one-loop two-point Green functions ${\cal G}_{ij}$. 
Here we use  ${\cal G}_{ij}$ to represent the full two-point Green functions, while the Gram matrix $G_{ij}$ denotes
 $x$-independent factor in the numerator of the two-point function, as given in (\ref{eq:def}).

Consider the one-loop correction of a two-point Green function ${\cal G}_{ij}^{(1)}$.
Its UV divergence is related to the one-loop renormalization matrix as
\begin{equation}
 - {\cal G}^{(1)}_{ij}\big|_{\mathrm{UV}}
= \langle \mathcal{O}_{i,\mathrm{b}}^\dag\,  (Z^{(1),k}_j \mathcal{O}_{k,\mathrm{b}}) \rangle
+\langle  (Z^{(1),k}_{i} \mathcal{O}_{k,\mathrm{b}})^\dag   \mathcal{O}_{j,\mathrm{b}} \rangle 
= Z^{(1),k}_j {\cal G}^{(0)}_{ik}
+Z^{(1),k}_{i} {\cal G}^{(0)}_{kj} \,.
\end{equation}

The second equality holds because $Z^{(1)}$ is a real matrix.
In terms of the Gram matrix and dilatation matrix, the above relation gives
\begin{equation}
\mathbb{D}^{(1),k}_j G^{(0)}_{ik}
+\mathbb{D}^{(1),k}_{i} G^{(0)}_{kj} 
= 2\tilde{\mathbb{D}}^{(1)}_{ij}\,.
\end{equation}
Here $\tilde{\mathbb{D}}^{(1)}_{ij}:= (\mathbb{D}^{(1)})_i^{\ k}G^{(0)}_{kj}$ is symmetric in $i,j$, which we will explain as follows.

The dilatation operation $\hat{\bm{D}}$ can be considered as a hermitian operator
acting on the quantum states which correspond to the gauge-invariant operators, 
so this gives $\langle i| (\hat{\bm{D}}j)\rangle=\langle (\hat{\bm{D}} i)  |j\rangle$.
Written in the matrix representation of $\hat{\bm{D}}$, the equation becomes
$\mathbb{D}^{(1),k}_{j}\langle i|k\rangle-
(\mathbb{D}^{(1),k}_{i})^{*}\langle k|j\rangle
=\mathbb{D}^{(1),k}_{j} G^{(0)}_{ik}-(\mathbb{D}^{(1),k}_{i})^{*}  G^{(0)}_{kj}=0$.
Recall that $G^{(0)}$ is symmetric and $\mathbb{D}^{(1)}$ is real, so 
finally one has $\mathbb{D}^{(1),k}_{j} G^{(0)}_{ki}- \mathbb{D}^{(1),k}_{i}   G^{(0)}_{kj}=0$.\footnote{In our computation, both $\mathbb{D}^{(1)}$ and $G^{(0)}$ is computed in $d=4-2\epsilon$. To check the symmetry of our result, one should deal with the epsilon expansion carefully. A more detailed discussion is given in Section~\ref{app:GZ symmetric}.}

Given the one-loop two-point functions, one can extract the matrix $\tilde{\mathbb{D}}^{(1)}$. 
In general the eigenvalues of $\tilde{\mathbb{D}}^{(1)}_{ij}$ are not equal to one-loop anomalous dimensions, since $\tilde{\mathbb{D}}^{(1)} \neq {\mathbb{D}}^{(1)}$.
To obtain the one-loop anomalous dimensions, one also needs the result of tree-level Gram matrix $G^{(0)}$ to get the dilation matrix.
In general $d$ dimensions,   $G^{(0)}$ is non-singular, so
\begin{equation}
{\mathbb{D}}^{(1)} = \tilde{\mathbb{D}}^{(1)} \cdot [G^{(0)}]^{-1} \,.
\end{equation}

Denote the rank of $G^{(0)}$ and $\mathbb{D}^{(1)}$ by $n$.
Consider an eigenvector of $\mathbb{D}^{(1)}$ denoted by $\mathbf{v}$, with corresponding eigenvalue $\lambda$ (which is an anomalous dimension):
\begin{align}
\mathbf{v}^{\mathrm{T}} \mathbb{D}^{(1)}=\lambda \mathbf{v}^{\mathrm{T}}\,.
\end{align}
For $\tilde{\mathbb{D}}^{(1)} =\mathbb{D}^{(1)}\cdot G^{(0)}$ it has
\begin{align}
\label{eq:eigen1}
\mathbf{v}^{\mathrm{T}} \tilde{\mathbb{D}}^{(1)} =\lambda \mathbf{v}^{\mathrm{T}} G^{(0)}\,.
\end{align}
It should be clear that in general the eigenvalues of $\tilde{\mathbb{D}}^{(1)}_{ij}$ are not
equal to one-loop anomalous dimensions $\lambda$.

\subsection{The signature of the Gram matrix and complex anomalous dimensions}

\label{sec:signature}

Below we prove that: for $\mathbb{D}^{(1)}$ to have complex eigenvalues, 
a necessary condition is that the Gram matrix $G^{(0)}$ must have both positive and negative eigenvalues. 
This statement is mentioned in Section~\ref{theadresults}.

Suppose $G^{(0)}$ and $\mathbb{D}^{(1)}$ have full rank $n$.
Since  $G^{(0)}$ is real and symmetric,  following the 
standard Gram-Schmidt algorithm \cite{cheney2009linear},
one can find a real matrix $\mathcal{N}$ such that
\begin{align}
G^{(0)}=\mathcal{N} \Lambda \mathcal{N}^{\mathrm{T}}\,,
\end{align}
where $\Lambda$ is the diagonal  matrix $\mathrm{diag}(1,\cdots,-1,\cdots)$,
and the numbers of $\pm 1$ equal the numbers of positive and negative eigenvalues of $G^{(0)}$.
Recall  $\tilde{\mathbb{D}}^{(1)} =\mathbb{D}^{(1)}\cdot G^{(0)}$ is real and symmetric.
Rotate $\tilde{\mathbb{D}}^{(1)}$ and $\mathbf{v}$ by $\mathcal{M}=\mathcal{N}\sqrt{\Lambda}$
(choose $\sqrt{-1}=\mathbbm{i}$),
so (\ref{eq:eigen1}) gives
\begin{align}
\label{eq:eigen2}
 \mathbf{u}_c^{\mathrm{T}}  \Theta_c=
\lambda \,\mathbf{u}_c^{\mathrm{T}}\,,
\end{align}
where
\begin{align}
\Theta_c:=\mathcal{M}^{-1}\tilde{\mathbb{D}}^{(1)}(\mathcal{M}^{-1})^{\mathrm{T}}\,,\quad
\mathbf{u}_c:=\mathcal{M}^{\mathrm{T}}\mathbf{v}\,.
\end{align}
The matrix $\Theta_c$ is symmetric and
shares the same eigenvalues with $\mathbb{D}^{(1)}$.
When the eigenvalues of $G^{(0)}$ are all positive or all negative then $\Theta_c$ is also real.
The eigenvalues of a real and symmetric matrix $\Theta_c$ must be real,
which proves the statement made in the beginning.

If $G^{(0)}$ have both positive and negative eigenvalues, 
$\Theta_c$ is a complex matrix, but its eigenvalues are not necessarily complex.
As a simple example, let's consider
\begin{align}
\mathbb{D}^{(1)}=\begin{pmatrix}
a_1+a_2 & -x\\x & a_1-a_2
\end{pmatrix}\,,\quad
G^{(0)}=\begin{pmatrix}
-1 & 0\\0 & 1
\end{pmatrix}\,,
\end{align}
where $a_1,a_2,x$ are all positive real numbers.
Correspondingly
\begin{align}
\tilde{\mathbb{D}}^{(1)}=\begin{pmatrix}
-a_1-a_2 & -x\\-x & a_1-a_2
\end{pmatrix}\,,\quad
\Theta_c=\begin{pmatrix}
a_1+a_2 & \mathbbm{i}x\\ \mathbbm{i} x& a_1-a_2
\end{pmatrix}\,.
\end{align}
The eigenvalues of this complex $\Theta_c$ are  $a_1\pm \sqrt{a_2^2-x^2}$.
The condition of eigenvalues being complex is that $|x|>|a_2|$.

\subsection{Symmetric condition on the dilatation matrix and the Gram matrix}
\label{app:GZ symmetric}

As mentioned in Section~\ref{sec:discuss},
the symmetric property of $\tilde{\mathbb{D}}^{(1)}=\mathbb{D}^{(1)}\cdot G^{(0)}$ provides 
a consistency check of our computation result.
When there exist evanescent operators, different blocks in $\tilde{\mathbb{D}}^{(1)}$ 
bear different $\epsilon$ orders, and therefore the equation
$\tilde{\mathbb{D}}^{(1)}=\tilde{\mathbb{D}}^{(1),\mathrm{T}}$ should be treated
more carefully.

Decompose $G^{(0)}$ and $\mathbb{D}^{(1)}$ in blocks of physical and evanescent sectors:
\begin{align}
\label{eq:GDeOrder}
\mathbb{D}^{(1)}=
\begin{pmatrix}
\mathbb{D}^{(1),\mathrm{p}}_{\mathrm{p}} &\mathbb{D}^{(1),\mathrm{e}}_{\mathrm{p}}\\
  \mathbb{D}^{(1),\mathrm{p}}_{\mathrm{e}} &  \mathbb{D}^{(1),\mathrm{e}}_{\mathrm{e}}
\end{pmatrix},\quad
G^{(0)}=
\begin{pmatrix}
G^{(0)}_{\mathrm{pp}} &   G^{(0)}_{\mathrm{pe}}\\
 G^{(0)}_{\mathrm{ep}} & G^{(0)}_{\mathrm{ee}}
\end{pmatrix}\,.
\end{align}
The subscripts p and e refer to `physical' and `evanescent'.
Expand all the blocks in $\epsilon$, \emph{e.g.} 
$G^{(0)}_{\mathrm{pp}}=\sum_{a=0} G^{(0)}_{\mathrm{pp},a} \,\epsilon^{a}$,
$\mathbb{D}^{(1),\mathrm{p}}_{\mathrm{p}}=\sum_{a=0} 
\mathbb{D}^{(1),\mathrm{p}}_{\mathrm{p},a} \,\epsilon^{a}$.
Notice that
$G^{(0)}_{\mathrm{ee},0}$, $G^{(0)}_{\mathrm{ep},0}$, $G^{(0)}_{\mathrm{pe},0}$
and $\mathbb{D}^{(1),\mathrm{p}}_{\mathrm{e},0}$ all vanish.

Write out the leading order of each block in $\tilde{\mathbb{D}}^{(1)}$:
\begin{align}
\label{check1}
\mathcal{O}(\epsilon^0):&\quad
\mathbb{D}^{(1),\mathrm{p}}_{\mathrm{p},0}
G^{(0)}_{\mathrm{pp},0}=\big(\mathbb{D}^{(1),\mathrm{p}}_{\mathrm{p},0}
G^{(0)}_{\mathrm{pp},0}\big)^{\mathrm{T}}\,,
\\
\label{check2}
\mathcal{O}(\epsilon^1):&\quad
 \mathbb{D}^{(1),\mathrm{e}}_{\mathrm{e},0} G^{(0)}_{\mathrm{ee},1}
=\big(\mathbb{D}^{(1),\mathrm{e}}_{\mathrm{e},0}
G^{(0)}_{\mathrm{ee},1}\big)^{\mathrm{T}}\,,
\\
\label{check3}
&\big(\mathbb{D}^{(1),\mathrm{e}}_{\mathrm{p},0} 
G^{(0)}_{\mathrm{ee},1}
+\mathbb{D}^{(1),\mathrm{p}}_{\mathrm{p},0} 
G^{(0)}_{\mathrm{pe},1} \big)^{\mathrm{T}}
=\mathbb{D}^{(1),\mathrm{e}}_{\mathrm{e},0} 
G^{(0)}_{\mathrm{ep},1}+\mathbb{D}^{(1),\mathrm{p}}_{\mathrm{e},1} 
G^{(0)}_{\mathrm{pp},0}\,.
\end{align}
These relations provide useful consistency checks for our results.

The above discussion is for a general $d$-dimensional theory. The 
$\mathbb{D}^{(1),\mathrm{p}}_{\mathrm{e},1}$ also appears when one performs renormalization in the finite renormalization scheme involving evanescent operators in a four-dimensional theory, see \cite{Buras:1989xd, Dugan:1990df, DiPietro:2017vsp, Jin:2023cce}. In that case, the dilatation one-loop matrix satisfies the relation~\eqref{check1}-\eqref{check3}.

\section{Explicit basis of Yang-Mills operators}
\label{app:opset}

In this section, we list the single-trace Yang-Mills operators
used in the above context.

The operators are grouped into
types of $C$-even and $C$-odd.
Within each $C$-parity type, the operators are
divided into physical and evanescent ones.
Finally, the physical operators are further classified
according to helicity sectors, which have been
introduced in Section~\ref{sec:positive}.

\subsection{Dimension 10}

\label{app:opdim10}

\subsubsection*{Length-3 operators }
\label{sec:dim10len3}

At mass dimension 10 and length 3, there are
five operators, four (one) of them are $C$-even (odd).
\begin{align}
\label{eq:dim10OpL3}
&\mathcal{O}_{10;3;\alpha;+;1}=
\frac{1}{12} (\partial^2) ^2 \text{tr}(F_{12}\, F_{13}\, F_{23})\,,
\nonumber\\
&\mathcal{O}_{10;3;\alpha;+;2}=\text{tr}(D_{12} F_{34}\, D_{12} F_{35}\, F_{45})\,,
\nonumber\\
&\mathcal{O}_{10;3;\beta;+;1}
=-\frac{1}{12} \partial^2  \Big(\partial^2  \text{tr}(F_{12}\, F_{13}\, F_{23})
-6 \text{tr}(D_1 F_{23}\,D_4 F_{23}\,F_{14})\Big)\,,
\nonumber\\
&\mathcal{O}_{10;3;\beta;+;2}
=\text{tr}(D_{12} F_{34}\,D_{15} F_{34}\,F_{25})-\text{tr}(D_{12} F_{34}\,D_{12} F_{35}\,F_{45})\,,
\nonumber\\
&\mathcal{O}_{10;3;\beta;-;1}
=\text{tr}(D_{12} F_{34}\,D_5 F_{34}\,D_1 F_{25})-\text{tr}(D_2 F_{34}\,D_{15} F_{34}\,D_1 F_{25})\,.
\end{align}
Their Gram matrix and one-loop dilatation matrix are given in 
(\ref{eq:G010L3str}) and (\ref{eq:Z110L3str}).

\subsubsection*{Length-4 operators}
\label{sec:dim10strEven}

At the level of dimension 10 and length 4, there are 24 independent single-trace operators, 
18   $C$-even and 6 $C$-odd.
All these operators are given in our previous work \cite{Jin:2022ivc},
and here we list the evanescent ones.

The three  $C$-even single-trace evanescent operators have been given
in  (\ref{eq:egforDtype}), and
their norms  are given in (\ref{eq:GofD10L4Eva}).
The $18\times 18$ Gram matrix of all the single-trace $C$-even
operators (15 physical and 3 evanescent) 
at the leading $N_c$ order is given in (\ref{eq:G0104s+}) .

The only $C$-odd evanescent single-trace operator is
\begin{align}
&\mathcal{O}^e_{10;4;\mathrm{s}-;1}=
\partial_{\mu}   \frac{1}{2}\delta^{1234 \mu}_{5678 9}
\text{tr}( D_9 F_{12} F_{34} F_{56} F_{78} )
-  \partial_{\mu}\partial_{\nu} \frac{1}{4}
\delta^{1234 \mu}_{5678 \nu}
\text{tr}( F_{12} F_{34}F_{78} F_{56} )
\,.
\end{align}
The $6\times 6$ Gram matrix of all the single-trace $C$-odd
operators (5 physical and 1 evanescent) 
at the leading $N_c$ order is given in (\ref{eq:G0104s-}) .

\subsubsection*{Length-5 single-trace operators}
\label{sec:dim10len5}

With mass dimension 10 and length 5,
there are six single-trace operators, all $C$-even.
Two out of them are evanescent operators,
and they are of $\delta$-5 type.
All these operators are given in our previous work \cite{Jin:2022ivc},
and here we list the evanescent ones.
\begin{align}
\label{eq:dim10OpL5}
& \mathcal{O}^e_{10;5;\mathrm{s}+;1}=\frac{1}{8}\delta^{12345}_{6789a}
\Big( \text{tr}( F_{12}  F_{67}  F_{34}  F_{89}  F_{5a} )
+\text{tr}( F_{12}  F_{67} F_{89}  F_{5a}  F_{34} )\Big)
\,,
\nonumber\\
& \mathcal{O}^e_{10;5;\mathrm{s}+;2}=\frac{1}{8}\delta^{12345}_{6789a}
\Big(-2 \text{tr}( F_{12}  F_{67}  F_{34}  F_{89}  F_{5a} )
+\text{tr}( F_{12}  F_{67} F_{89}  F_{5a}  F_{34} )\Big)
\,.
\end{align}
Their Gram matrix and one-loop dilatation matrix are given in 
(\ref{eq:G0105s}) and (\ref{eq:ZD10L5}).

\subsection{Dimension 12}

\label{app:opdim12}

\subsubsection*{Dim-12 length-4 evanescent operators}

At the level of dimension 12 and length 4, there are 107 independent single-trace operators,
68 $C$-even and 39 $C$-odd.
All these operators are given in the ancillary files, and
here we list all the evanescent operators, including
16 $C$-even and 9 $C$-odd ones.

The $C$-even single-trace sector can be divided into 6 $D$-sectors as below. The $D$-(4,2) sector reads
\begin{equation}
\begin{aligned}
&\partial^2 \partial_9 \partial_\mu\Big[\delta^{12349}_{5678\mu }\Big(\text{tr}(F_{12}F_{34}F_{56}F_{78})+\text{trace reverse})\Big)\Big],\\
&\partial^2 \partial_9 \partial_\mu\Big[\delta^{12359}_{4678\mu }\Big(\text{tr}(F_{12}F_{34}F_{56}F_{78})+\text{trace reverse})\Big)\Big]\,.
\end{aligned}
\end{equation}
The $D$-(3,3) sector reads
\begin{align}
	&\partial_\mu \partial_\nu \partial_\rho\Big[\delta^{1256\nu }_{3789\rho } \delta^{4}_{\mu }\Big(\text{tr}(D_{1}F_{23}F_{45}F_{67}F_{89})+\text{trace reverse})\Big)\Big]\,.
\end{align}
The $D$-(2,0) sector reads
\begin{align}
	&\partial^2\Big[\delta^{12456}_{3789\mu }\Big(\text{tr}(D_{1}F_{23}F_{45}D_{6}F_{78}F_{9\mu })+\text{trace reverse})\Big)\Big]\,.
\end{align}
The $D$-(2,2) sector reads
\begin{equation}
\begin{aligned}
	&\partial_\nu \partial_\rho\Big[\delta^{12456\nu }_{3789\mu \rho }\Big(\text{tr}(D_{1}F_{23}F_{45}D_{6}F_{78}F_{9\mu })+\text{trace reverse})\Big)\Big],\\
	&\partial_\nu \partial_\rho\Big[\delta^{1}_{4} \delta^{2356\nu }_{789\mu \rho }\Big(\text{tr}(D_{1}F_{23}F_{45}D_{6}F_{78}F_{9\mu })+\text{trace reverse})\Big)\Big],\\
	&\partial_\nu \partial_\rho\Big[\delta^{1}_{4} \delta^{2356\nu }_{789\mu \rho }\Big(\text{tr}(D_{1}F_{23}D_{4}F_{56}F_{78}F_{9\mu })+\text{trace reverse})\Big)\Big],\\
	&\partial_\nu \partial_\rho\Big[\delta^{1}_{4} \delta^{2357\nu }_{689\mu \rho }\Big(\text{tr}(D_{1}F_{23}D_{4}F_{56}F_{78}F_{9\mu })+\text{trace reverse})\Big)\Big],\\
	&\partial_\nu \partial_\rho\Big[\delta^{1}_{4} \delta^{2367\nu }_{589\mu \rho }\Big(\text{tr}(D_{1}F_{23}F_{45}D_{6}F_{78}F_{9\mu })+\text{trace reverse})\Big)\Big],\\
	&\partial_\nu \partial_\rho\Big[\delta^{1}_{4} \delta^{2378\nu }_{569\mu \rho }\Big(\text{tr}(D_{1}F_{23}D_{4}F_{56}F_{78}F_{9\mu })+\text{trace reverse})\Big)\Big],\\
	&\partial_\nu \partial_\rho\Big[\delta^{1}_{5} \delta^{2347\nu }_{689\mu \rho }\Big(\text{tr}(D_{1}F_{23}D_{4}F_{56}F_{78}F_{9\mu })+\text{trace reverse})\Big)\Big],\\
	&\partial_\nu \partial_\rho\Big[\delta^{2}_{4}\delta^{1567\nu }_{389\mu \rho } \Big(\text{tr}(D_{1}F_{23}F_{45}D_{6}F_{78}F_{9\mu })+\text{trace reverse})\Big)\Big]\,.	
\end{aligned}
\end{equation}
The $D$-(2,2) sector is also shown in (\ref{delta6D22})-(\ref{lastD22}). The 
one-loop dilatation matrix of this sector is given in (\ref{D22length4talbe}).
The norm of the only $\delta_6$ operator is given in (\ref{eq:opd12delta6b}).

The $D$-(1,1) sector reads
\begin{equation}
\begin{aligned}
	&\partial_\rho\Big[\delta^{1}_{4} \delta^{23578}_{69\mu \nu \rho }\Big(\text{tr}(D_{1}F_{23}D_{4}F_{56}D_{7}F_{89}F_{\mu \nu })+\text{trace reverse})\Big)\Big],\\
	&\partial_\rho\Big[\delta^{1}_{5} \delta^{23478}_{69\mu \nu \rho }\Big(\text{tr}(D_{1}F_{23}D_{4}F_{56}D_{7}F_{89}F_{\mu \nu })+\text{trace reverse})\Big)\Big]\,.
\end{aligned}
\end{equation}
The $D$-(0,0) sector reads
\begin{equation}
\begin{aligned}
	&\delta^{1}_{4} \delta^{2357\mu }_{689\nu \rho }\Big(\text{tr}(D_{1}F_{23}D_{4}F_{56}D_{7}F_{89}D_{\mu }F_{\nu \rho })+\text{trace reverse})\Big),\\&\delta^{1}_{4} \delta^{2378\mu }_{569\nu \rho }\Big(\text{tr}(D_{1}F_{23}D_{4}F_{56}D_{7}F_{89}D_{\mu }F_{\nu \rho })+\text{trace reverse})\Big)\,.
\end{aligned}
\end{equation}
There are 4 $D$-sectors in the C-odd sector. The $D$-(3,1) sector reads
\begin{align}
	&\partial^2 \partial_\mu\Big[\delta^{12456}_{3789\mu }\Big(\text{tr}(D_{1}F_{23}F_{45}F_{67}F_{89})-\text{trace reverse})\Big)\Big]\,.
\end{align}
The $D$-{3,3} sector reads
\begin{equation}
\begin{aligned}
	&\partial_\mu \partial_\nu \partial_\rho\Big[\delta^{1}_{\mu } \delta^{2345\nu }_{6789\rho }\Big(\text{tr}(D_{1}F_{23}F_{45}F_{67}F_{89})-\text{trace reverse})\Big)\Big],\\
	&\partial_\mu \partial_\nu \partial_\rho\Big[\delta^{2}_{\mu }\delta^{1456\nu }_{3789\rho } \Big(\text{tr}(D_{1}F_{23}F_{45}F_{67}F_{89})-\text{trace reverse})\Big)\Big]\,.
\end{aligned}
\end{equation}
The $D$-(2,2) sector reads
\begin{equation}
\begin{aligned}
	&\partial_\nu \partial_\rho\Big[\delta^{1456\nu }_{3789\rho } \delta^{2}_{\mu }\Big(\text{tr}(D_{1}F_{23}F_{45}F_{67}F_{89})-\text{trace reverse})\Big)\Big],\\
	&\partial_\nu \partial_\rho\Big[\delta^{1}_{4} \delta^{2356\nu }_{789\mu \rho }\Big(\text{tr}(D_{1}F_{23}F_{45}D_{6}F_{78}F_{9\mu })-\text{trace reverse})\Big)\Big],\\
	&\partial_\nu \partial_\rho\Big[\delta^{1}_{4} \delta^{2367\nu }_{589\mu \rho }\Big(\text{tr}(D_{1}F_{23}F_{45}D_{6}F_{78}F_{9\mu })-\text{trace reverse})\Big)\Big]\,.
\end{aligned}
\end{equation}
The $D$-(1,1) sector reads
\begin{equation}
\begin{aligned}
	&\partial_\rho\Big[\delta^{1}_{4} \delta^{23567}_{89\mu \nu \rho }\Big(\text{tr}(D_{1}F_{23}D_{4}F_{56}D_{7}F_{89}F_{\mu \nu })-\text{trace reverse})\Big)\Big],\\
	&\partial_\rho\Big[\delta^{1}_{4} \delta^{23578}_{69\mu \nu \rho }\Big(\text{tr}(D_{1}F_{23}D_{4}F_{56}D_{7}F_{89}F_{\mu \nu })-\text{trace reverse})\Big)\Big],\\
	&\partial_\rho\Big[\delta^{1}_{5} \delta^{23478}_{69\mu \nu \rho }\Big(\text{tr}(D_{1}F_{23}D_{4}F_{56}D_{7}F_{89}F_{\mu \nu })-\text{trace reverse})\Big)\Big]\,.
\end{aligned}
\end{equation}

\subsubsection*{Dim-12 length-5 evanescent operators of $\delta_6$ type }

At the level of dimension 12 and length 5, there are 151 independent single-trace operators,
92 $C$-even and 59 $C$-odd. 
Among them, there are 61 evanescent operators, 
36 $C$-even and 25 $C$-odd.

All these operators are given in the ancillary files, and
here we list all the 8 $\delta_6$ operators, 6 $C$-even
and 2 $C$-odd. 

The two in the $C$-even $D$-(2,2) sector read
\begin{equation}
\begin{aligned}
&\partial_\nu \partial_\rho\Big[\delta^{12345\nu }_{6789\mu \rho }\Big(\text{tr}(F_{12}F_{34}F_{56}F_{78}F_{9\mu })-\text{trace reverse})\Big)\Big],\label{eq:dim12lengh5delta6-1}\\
&\partial_\nu \partial_\rho\Big[\delta^{12357\nu }_{4689\mu \rho }\Big(\text{tr}(F_{12}F_{34}F_{56}F_{78}F_{9\mu })-\text{trace reverse})\Big)\Big]\,.
\end{aligned}
\end{equation}
The two in the $C$-even $D$-(1,1) sector read
\begin{equation}
\begin{aligned}
	&\partial_\rho\Big[\delta^{124567}_{389\mu \nu \rho }\Big(\text{tr}(D_{1}F_{23}F_{45}F_{67}F_{89}F_{\mu \nu })-\text{trace reverse})\Big)\Big],\\
	&\partial_\rho\Big[\delta^{12456\mu }_{3789\nu \rho }\Big(\text{tr}(D_{1}F_{23}F_{45}F_{67}F_{89}F_{\mu \nu })-\text{trace reverse})\Big)\Big]\,.
\end{aligned}
\end{equation}
The two in the $C$-even $D$-(0,0) sector read
\begin{equation}
\begin{aligned}
	&\delta^{124567}_{389\mu \nu \rho }\Big(\text{tr}(D_{1}F_{23}F_{45}D_{6}F_{78}F_{9\mu }F_{\nu \rho })-\text{trace reverse})\Big),\\&\delta^{124568}_{379\mu \nu \rho }\Big(\text{tr}(D_{1}F_{23}F_{45}F_{67}D_{8}F_{9\mu }F_{\nu \rho })-\text{trace reverse})\Big)\,.
\end{aligned}
\end{equation}
The two in the $C$-odd $D$-(1,1) sector read
\begin{equation}
\begin{aligned}
	&\partial_\rho\Big[\delta^{124567}_{389\mu \nu \rho }\Big(\text{tr}(D_{1}F_{23}F_{45}F_{67}F_{89}F_{\mu \nu })+\text{trace reverse})\Big)\Big],\\
	&\partial_\rho\Big[\delta^{124568}_{379\mu \nu \rho }\Big(\text{tr}(D_{1}F_{23}F_{45}F_{67}F_{89}F_{\mu \nu })+\text{trace reverse})\Big)\Big]\,.\label{eq:dim12lengh5delta6-2}
\end{aligned}
\end{equation}


\section{The Gram matrices and dilatation matrices} 
\label{ap:otherG}

In this section, we give the Gram matrices and dilation matrices not shown above.

\subsection{Dimension 8}
The Gram matrices of operators (\ref{eq:dim8L3oper}) and (\ref{eq:dim8L4stroper})
has been given in (\ref{eq:G083s}) and (\ref{eq:G084s}), and the corresponding
one-loop dilatation matrices are
\begin{align}
	&\mathbbm{D}^{(1)}_{8;3}= N_c 
	\left(
		\begin{array}{c;{2pt/2pt}c }
	   \frac{7}{3}  $\,$  & $\,$ 0    \\[0.2em]
	    \hdashline[2pt/2pt]\rule{0pt}{0.9\normalbaselineskip}
	      0   $\,$   &  $\,$  1   \\
		\end{array}
		\right)\,,
\\
	&\mathbbm{D}^{(1)}_{8;4 }=N_c
	2\left(
	\begin{array}{cc;{2pt/2pt}cc}
		-\frac{26 }{3} & 2 & 0 & 0
		\\[0.2em]
		-4 & \frac{16}{2} & 0 &  0
		\\[0.2em]
		\hdashline[2pt/2pt]\rule{0pt}{0.9\normalbaselineskip}
		0 & 0 & -\frac{19}{3} & -1
		\\[0.2em]
		0 & 0 & -2 & 2
	\end{array}
	\right)\,.
\end{align}

\subsection{Dimension 10}\label{app:gramdim10}

\subsubsection*{Length 3}
Among five dim-10 length-3 operators given in (\ref{eq:dim10OpL3}),
four (one) are $C$-even (odd).
Operators with different $C$-parities are orthogonal,
so their Gram matrices can be written individually.
At the leading order of $N_c$ they read
\begin{align}
	\label{eq:G010L3str}
	&G _{10;3;+}=2\,N_c^3\, (d-2)^4(d-1)d^2(d+1)(d+2)^2(d+3)(d+4)
	\nonumber\\
	&{\small\times
		\left(
		\begin{array}{cc;{2pt/2pt}cc}
			3 (3 d-8) (3 d+2) (3 d+4) & 3 (d+2) (3 d-8) (3 d+4) & -3 (d-4) (3 d+2) (3 d+4) & -3 (d-4) (d+2) (3 d+4) \\[0.2em]
			3 (d+2) (3 d-8) (3 d+4) & (3 d-8) \left(3 d^2+14 d+24\right) & -3 (d-4) (d+2) (3 d+4) & -(d-4) \left(3 d^2+14 d+24\right) \\[0.2em]
			\hdashline[2pt/2pt]\rule{0pt}{0.9\normalbaselineskip}
			-3 (d-4) (3 d+2) (3 d+4) & -3 (d-4) (d+2) (3 d+4) & 3 (d-2) (d+8) (3 d+4) & 3 (d+2) \left(d^2+12 d-32\right) \\[0.2em]
			-3 (d-4) (d+2) (3 d+4) & -(d-4) \left(3 d^2+14 d+24\right) & 3 (d+2) \left(d^2+12 d-32\right) & d^3+26 d^2+8 d-192 \\[0.2em]
		\end{array}
		\right)\,,}
	\nonumber\\
	&G _{10;3;-}=32\,N_c^3\,  (d-3)(d-2)^4 (d-1)d^3(d+1)(d+2)^2(d+3)(d+4)
	\,.
\end{align}
Still, the vertical and horizontal dashed lines divide the helicity sectors.
The helicity-crossing matrix elements vanish at $d=4$  as expected.

The dilatation matrix of these five operators is
\begin{align}
\label{eq:Z110L3str}
\mathbb{D}^{(1)}_{10;3;+}= N_c 
	\left(
		\begin{array}{cc;{2pt/2pt}cc}
	   $\,\,$ \frac{7}{3} & 0 $\,$ & 0 & 0 \\
	    -\frac{6}{5} $\,\,$ & \frac{71}{15}  $\,$ & 0 & 0 \\[0.3em]
	    \hdashline[2pt/2pt]\rule{0pt}{0.9\normalbaselineskip}
	    $\,\,$  0 & 0 $\,$ & 1 & 0 \\
	   $\,\,$   0 & 0 $\,$ & -2 & \frac{17}{3}
		\end{array}
		\right)\,,\quad
\mathbb{D}^{(1)}_{10;3;-}= \frac{13}{3}N_c\,.
\end{align}

\subsubsection*{Length 4}
\label{app:GZ-D10L4}

Here we give the 
Gram matrix of dim-10 length-4 $C$-even and $C$-odd single-trace operators at
leading $N_c$ order, whose property has been summarized
in  Section~\ref{sec:positive}. The corresponding operators are given in our
previous work \cite{Jin:2022ivc}, and also in the ancillary file of this paper.

The $18\times 18$ leading-color Gram matrix of   $C$-even sector is
\begin{align}
	\label{eq:G0104s+}
	G _{10;4;+}=512 N_c^4 (d-2)^5d^2(d-1)
	\times
	\left(
	\begin{array}{c ;{2pt/2pt}c }
		M_1 & M_2
	\end{array}
	\right)+\mathcal{O}(\epsilon^2)\,,
\end{align}
where
\begin{footnotesize}
	\begin{align}
		&M_1=
		\nonumber\\
		&\left(
		\begin{array}{ccccc;{2pt/2pt}cccc}
			\frac{945}{2}-\frac{32619 \epsilon }{16} & 126 \epsilon  & \frac{30099 \epsilon }{16}-\frac{945}{2} & -\frac{243 \epsilon }{2}
			& \frac{14391 \epsilon }{16}-\frac{405}{2} & 0 & -315 \epsilon  & -\frac{315 \epsilon }{2} & 45 \epsilon  \\[0.2em]
			126 \epsilon  & 6048-26196 \epsilon  & 126 \epsilon  & 2352 \epsilon -432 & -396 \epsilon  & 756 \epsilon  & 0 & 0 & -216
			\epsilon  \\[0.2em]
			\frac{30099 \epsilon }{16}-\frac{945}{2} & 126 \epsilon  & 495-\frac{14959 \epsilon }{8} & \frac{167 \epsilon }{4} &
			225-\frac{7135 \epsilon }{8} & 0 & 385 \epsilon  & 145 \epsilon  & -20 \epsilon  \\[0.2em]
			-\frac{243 \epsilon }{2} & 2352 \epsilon -432 & \frac{167 \epsilon }{4} & \frac{63}{2}-\frac{1073 \epsilon }{4} & \frac{199
				\epsilon }{2} & -60 \epsilon  & 77 \epsilon  & -10 \epsilon  & 35 \epsilon  \\[0.2em]
			\frac{14391 \epsilon }{16}-\frac{405}{2} & -396 \epsilon  & 225-\frac{7135 \epsilon }{8} & \frac{199 \epsilon }{2} &
			\frac{225}{2}-\frac{3985 \epsilon }{8} & 0 & 150 \epsilon  & 80 \epsilon  & -25 \epsilon  \\[0.2em]
			\hdashline[2pt/2pt]\rule{0pt}{0.9\normalbaselineskip}
			0 & 756 \epsilon  & 0 & -60 \epsilon  & 0 & 420-\frac{2831 \epsilon }{2} & \frac{980 \epsilon }{3} & \frac{280 \epsilon }{3} &
			-\frac{140 \epsilon }{3} \\[0.2em]
			-315 \epsilon  & 0 & 385 \epsilon  & 77 \epsilon  & 150 \epsilon  & \frac{980 \epsilon }{3} & \frac{2800}{3}-\frac{29290
				\epsilon }{9} & \frac{350}{3}-\frac{16535 \epsilon }{36} & \frac{2131 \epsilon }{9}-\frac{280}{3} \\[0.2em]
			-\frac{315 \epsilon }{2} & 0 & 145 \epsilon  & -10 \epsilon  & 80 \epsilon  & \frac{280 \epsilon }{3} &
			\frac{350}{3}-\frac{16535 \epsilon }{36} & \frac{155}{6}-\frac{17803 \epsilon }{144} & \frac{10}{3}-\frac{253 \epsilon }{36}
			\\[0.2em]
			45 \epsilon  & -216 \epsilon  & -20 \epsilon  & 35 \epsilon  & -25 \epsilon  & -\frac{140 \epsilon }{3} & \frac{2131 \epsilon
			}{9}-\frac{280}{3} & \frac{10}{3}-\frac{253 \epsilon }{36} & \frac{100}{3}-\frac{955 \epsilon }{9} \\[0.2em]
			\hdashline[2pt/2pt]\rule{0pt}{0.9\normalbaselineskip}
			-567 \epsilon  & 756 \epsilon  & 378 \epsilon  & -216 \epsilon  & 270 \epsilon  & 378 \epsilon  & 504 \epsilon  & 0 & 72
			\epsilon  \\[0.2em]
			-63 \epsilon  & -2772 \epsilon  & 63 \epsilon  & 279 \epsilon  & -54 \epsilon  & -378 \epsilon  & -504 \epsilon  & 0 & 108
			\epsilon  \\[0.2em]
			-63 \epsilon  & -1260 \epsilon  & \frac{329 \epsilon }{2} & \frac{329 \epsilon }{2} & 63 \epsilon  & -56 \epsilon  &
			-\frac{2702 \epsilon }{3} & -\frac{490 \epsilon }{3} & \frac{8 \epsilon }{3} \\[0.2em]
			9 \epsilon  & 450 \epsilon  & -\frac{43 \epsilon }{4} & -\frac{97 \epsilon }{2} & 11 \epsilon  & 60 \epsilon  & 79 \epsilon  &
			0 & -15 \epsilon  \\[0.2em]
			-\frac{405 \epsilon }{2} & 216 \epsilon  & 149 \epsilon  & -\frac{267 \epsilon }{4} & 100 \epsilon  & 140 \epsilon  & \frac{464
				\epsilon }{3} & -\frac{5 \epsilon }{3} & \frac{70 \epsilon }{3} \\[0.2em]
			135 \epsilon  & -702 \epsilon  & -\frac{211 \epsilon }{4} & \frac{589 \epsilon }{4} & -\frac{433 \epsilon }{4} & 0 & -189
			\epsilon  & 10 \epsilon  & -20 \epsilon  \\[0.2em]
			\hline\rule{0pt}{0.9\normalbaselineskip}
			-1260 \epsilon  & 5040 \epsilon  & 700 \epsilon  & -976 \epsilon  & 820 \epsilon  & 280 \epsilon  & \frac{1120 \epsilon }{3} &
			-\frac{700 \epsilon }{3} & \frac{560 \epsilon }{3} \\[0.2em]
			-1260 \epsilon  & -4032 \epsilon  & 910 \epsilon  & 218 \epsilon  & 400 \epsilon  & 280 \epsilon  & -\frac{1400 \epsilon }{3} &
			-\frac{700 \epsilon }{3} & \frac{560 \epsilon }{3} \\[0.2em]
			810 \epsilon  & 2592 \epsilon  & -630 \epsilon  & -168 \epsilon  & -270 \epsilon  & -280 \epsilon  & \frac{1400 \epsilon }{3} &
			\frac{580 \epsilon }{3} & -\frac{320 \epsilon }{3} \\[0.2em]
		\end{array}
		\right)\,,
	\end{align}
\end{footnotesize}
and
\begin{footnotesize}
	\begin{align}
		&M_2=
		\nonumber\\
		&\left(
		\begin{array}{cccccc|ccc}
			-567 \epsilon  & -63 \epsilon  & -63 \epsilon  & 9 \epsilon  & -\frac{405 \epsilon }{2} & 135 \epsilon  & -1260 \epsilon  &
			-1260 \epsilon  & 810 \epsilon  \\[0.2em]
			756 \epsilon  & -2772 \epsilon  & -1260 \epsilon  & 450 \epsilon  & 216 \epsilon  & -702 \epsilon  & 5040 \epsilon  & -4032
			\epsilon  & 2592 \epsilon  \\[0.2em]
			378 \epsilon  & 63 \epsilon  & \frac{329 \epsilon }{2} & -\frac{43 \epsilon }{4} & 149 \epsilon  & -\frac{211 \epsilon }{4} &
			700 \epsilon  & 910 \epsilon  & -630 \epsilon  \\[0.2em]
			-216 \epsilon  & 279 \epsilon  & \frac{329 \epsilon }{2} & -\frac{97 \epsilon }{2} & -\frac{267 \epsilon }{4} & \frac{589
				\epsilon }{4} & -976 \epsilon  & 218 \epsilon  & -168 \epsilon  \\[0.2em]
			270 \epsilon  & -54 \epsilon  & 63 \epsilon  & 11 \epsilon  & 100 \epsilon  & -\frac{433 \epsilon }{4} & 820 \epsilon  & 400
			\epsilon  & -270 \epsilon  \\[0.2em]
			\hdashline[2pt/2pt]\rule{0pt}{0.9\normalbaselineskip}
			378 \epsilon  & -378 \epsilon  & -56 \epsilon  & 60 \epsilon  & 140 \epsilon  & 0 & 280 \epsilon  & 280 \epsilon  & -280
			\epsilon  \\[0.2em]
			504 \epsilon  & -504 \epsilon  & -\frac{2702 \epsilon }{3} & 79 \epsilon  & \frac{464 \epsilon }{3} & -189 \epsilon  &
			\frac{1120 \epsilon }{3} & -\frac{1400 \epsilon }{3} & \frac{1400 \epsilon }{3} \\[0.2em]
			0 & 0 & -\frac{490 \epsilon }{3} & 0 & -\frac{5 \epsilon }{3} & 10 \epsilon  & -\frac{700 \epsilon }{3} & -\frac{700 \epsilon
			}{3} & \frac{580 \epsilon }{3} \\[0.2em]
			72 \epsilon  & 108 \epsilon  & \frac{8 \epsilon }{3} & -15 \epsilon  & \frac{70 \epsilon }{3} & -20 \epsilon  & \frac{560
				\epsilon }{3} & \frac{560 \epsilon }{3} & -\frac{320 \epsilon }{3} \\[0.2em]
			\hdashline[2pt/2pt]\rule{0pt}{0.9\normalbaselineskip}
			756-\frac{6171 \epsilon }{2} & -378 \epsilon  & \frac{5415 \epsilon }{2}-756 & 54 \epsilon  & 270-\frac{4269 \epsilon }{4} &
			270 \epsilon  & -1512 \epsilon  & -1512 \epsilon  & 864 \epsilon  \\[0.2em]
			-378 \epsilon  & 1512-6549 \epsilon  & 378 \epsilon  & 978 \epsilon -216 & -108 \epsilon  & -135 \epsilon  & 1512 \epsilon  &
			-3024 \epsilon  & 1728 \epsilon  \\[0.2em]
			\frac{5415 \epsilon }{2}-756 & 378 \epsilon  & 1036-\frac{7643 \epsilon }{2} & -\frac{101 \epsilon }{2} & \frac{3443 \epsilon
			}{4}-242 & -\frac{309 \epsilon }{2} & 392 \epsilon  & -28 \epsilon  & 172 \epsilon  \\[0.2em]
			54 \epsilon  & 978 \epsilon -216 & -\frac{101 \epsilon }{2} & \frac{63}{2}-\frac{597 \epsilon }{4} & \frac{63 \epsilon }{4} &
			\frac{105 \epsilon }{4} & -272 \epsilon  & 502 \epsilon  & -288 \epsilon  \\[0.2em]
			270-\frac{4269 \epsilon }{4} & -108 \epsilon  & \frac{3443 \epsilon }{4}-242 & \frac{63 \epsilon }{4} & 100-\frac{3047 \epsilon
			}{8} & 85 \epsilon  & -484 \epsilon  & -484 \epsilon  & 280 \epsilon  \\[0.2em]
			270 \epsilon  & -135 \epsilon  & -\frac{309 \epsilon }{2} & \frac{105 \epsilon }{4} & 85 \epsilon  & 27-261 \epsilon  & 952
			\epsilon  & 70 \epsilon  & -6 \epsilon  \\[0.2em]
			\hline\rule{0pt}{0.9\normalbaselineskip}
			-1512 \epsilon  & 1512 \epsilon  & 392 \epsilon  & -272 \epsilon  & -484 \epsilon  & 952 \epsilon  & -6720 \epsilon  & 0 & 0 \\[0.2em]
			-1512 \epsilon  & -3024 \epsilon  & -28 \epsilon  & 502 \epsilon  & -484 \epsilon  & 70 \epsilon  & 0 & -8400 \epsilon  & 5040
			\epsilon  \\[0.2em]
			864 \epsilon  & 1728 \epsilon  & 172 \epsilon  & -288 \epsilon  & 280 \epsilon  & -6 \epsilon  & 0 & 5040 \epsilon  & -3120
			\epsilon  \\[0.2em]
		\end{array}
		\right)\,.
	\end{align}
\end{footnotesize}

The $6\times 6$ leading-color Gram matrix of $C$-odd sector is
\begin{align}
	\label{eq:G0104s-}
	& G _{10;4;-} =
	128 N_c^4 (d-2)^5d^2(d-1)  
	\nonumber\\
	&\times\Bigg[\left(
	\begin{array}{c;{2pt/2pt}cc;{2pt/2pt}cc|c}
		-2 (151 \epsilon -40) & 0 & 0 & -80 \epsilon  & -64 \epsilon  & -160 \epsilon  \\[0.2em]
		\hdashline[2pt/2pt]\rule{0pt}{0.9\normalbaselineskip}
		0 & -2 (1317 \epsilon -280) & - 1177 \epsilon +280 & 112 \epsilon  & 112 \epsilon  & 1120 \epsilon
		\\[0.2em]
		0 & -1177 \epsilon+280 & -5 (159 \epsilon -40) & 0 & 80 \epsilon  & 320 \epsilon  \\[0.2em]
		\hdashline[2pt/2pt]\rule{0pt}{0.9\normalbaselineskip}
		-80 \epsilon  & 112 \epsilon  & 0 & -4 (101 \epsilon -24) & 0 & -288 \epsilon  \\[0.2em]
		-64 \epsilon  & 112 \epsilon  & 80 \epsilon  & 0 & -6 (97 \epsilon -24) & -288 \epsilon  \\[0.2em]
		\hline\rule{0pt}{0.9\normalbaselineskip}
		-160 \epsilon  & 1120 \epsilon  & 320 \epsilon  & -288 \epsilon  & -288 \epsilon  & -1920 \epsilon
		\\
	\end{array}
	\right)
	+ \mathcal{O}(\epsilon^2)\Bigg]\,.
\end{align}
The solid lines are added to divide physical and evanescent operators,
while the dashed lines are added to divide different helicity sectors within physical operators.
Here $d=4-2\epsilon$  and we only keep the result up to $\mathcal{O}(\epsilon^1)$.
In the limit of $\epsilon\rightarrow 0$,
one can see the orthogonality between different helicity sectors, and also the
vanishing rows and columns corresponding to evanescent operators.
The chosen basis operators of each helicity sector
are
also classified into different $D$-types, and as we see different
$D$-types are not orthogonal to each other.
The full $N_c$-dependence of $G^{(0)}_{10;4;+}$, which is not given explicitly above,
is much more complicated than the overall factor in $G^{(0)}_{10;4;-}$,
and this is because $C$-even single-trace operators also mix with 
double-trace ones, while $C$-single operators do not.

Both (\ref{eq:G0104s+}) and (\ref{eq:G0104s-}) are positive definite when 
$\epsilon<0$ ($4<d<5$).
As explained in Appendix~\ref{sec:signature}, 
the coexistence of positive and negative eigenvalues of
the Gram matrix is the necessary condition for the existence of 
complex ADs. 
One can predict that all the ADs from dim-10 length-4 operators are real numbers.
This is confirmed by the concrete calculation.
It is sufficient to only inspect the ADs from evanescent sectors since
physical operators never create complex ADs.
The one-loop dilatation matrix of dim-10 length-4 operators has 
been given in our previous work \cite{Jin:2022ivc}, and here we rewrite the blocks
of single-trace evanescent operators:
\begin{align}
\label{eq:fullColorZD10L4}
	&\mathbbm{D}^{(1),\mathrm{e}}_{10;4;+}=N_c
	\left(
	\begin{array}{ccc}
		-\frac{4}{3} & -\frac{16}{3} & 0 \\[0.2em]
		-\frac{20}{3} & $\,$2 & 0\\
		$\,\,$4 & $\,$2 & \frac{16}{3}
	\end{array}
	\right)\,,
	\\
	&\mathbbm{D}^{(1),\mathrm{e}}_{10;4;-}= \frac{8 N_c}{3}\,. 
\end{align}
The anomalous dimensions from dim-10 length-4 evanescent operators are all real.

\subsubsection*{Length 5}

At $d=4-2\epsilon$, 
we give the leading-color Gram matrix of dim-10 length-5 single-trace operators.
We expand $d$ in $\epsilon$ and keep the order  up to $\mathcal{O}(\epsilon)$.
At the leading order of $N_c$, 
the Gram matrix of single-trace operators (given in our
previous work \cite{Jin:2022ivc}, and also in the ancillary file of this paper) is
\begin{align}
	\label{eq:G0105s}
	&G _{10;5} =32N_c^5 (d-2)^6d (d-1)
	\nonumber\\
	&\times\Bigg[\left(
	\begin{array}{cc;{2pt/2pt}cc|cc}
		200-960 \epsilon  & 140-640 \epsilon  & -90 \epsilon  & -60 \epsilon  & -160 \epsilon  & 140
		\epsilon  \\[0.2em]
		140-640\epsilon  & 100-445 \epsilon  & -45 \epsilon  & -38 \epsilon  & -72 \epsilon  & 30 
		\epsilon \\[0.2em]
		\hdashline[2pt/2pt]\rule{0pt}{0.9\normalbaselineskip}
		-90 \epsilon  & -45 \epsilon  & 12-73 \epsilon  & -18 \epsilon  & -72 \epsilon  & 90 
		\epsilon  \\[0.2em]
		-60 \epsilon  & -38 \epsilon  & -18 \epsilon  & 48-196 \epsilon  & -48 \epsilon  & 60 
		\epsilon  \\[0.2em]
		\hline\rule{0pt}{0.9\normalbaselineskip}
		-160 \epsilon  & -72 \epsilon  & -72 \epsilon  & -48 \epsilon  & -240 \epsilon  & 120 
		\epsilon  \\[0.2em]
		140 \epsilon  & 30 \epsilon  & 90 \epsilon  & 60 \epsilon  & 120 \epsilon  & -600 \epsilon  
		\\[0.2em]
	\end{array}
	\right)
	+\mathcal{O}(\epsilon^2)\Bigg]\,.
\end{align}
Still, the solid lines  divide physical and evanescent operators,
while the dashed lines divide helicity sectors in physical operators.
Similar to the length-4 cases, in the limit of $\epsilon\rightarrow 0$,
one can see the orthogonality between different helicity sectors, and also the
vanishing rows and columns corresponding to evanescent operators.

When $\epsilon<0$ ($4<d<5$), $G_{10;5;s}$ is positive definite, and
according to Appendix~\ref{sec:signature},
this predicts the absence of complex ADs.
The one-loop dilatation matrix of dim-10 length-5 operators has been 
given in our previous work \cite{Jin:2022ivc}, and here we rewrite the result
of single-trace operators,
	\begin{align}
	\label{eq:ZD10L5}
		&\mathbbm{D}^{(1)}_{10;5}=N_c
		\left(
	\begin{array}{cc;{2pt/2pt}cc|cc}
		-\frac{55}{3}  & 20  & 0 & 0  & \frac{25}{9} & -\frac{55}{9}   \\[0.2em]
		-18  & \frac{65}{3}  & 0 & 0  & \frac{20}{9} & -\frac{38}{9}   \\[0.2em]
		\hdashline[2pt/2pt]\rule{0pt}{0.9\normalbaselineskip}
		0  & 0 & -2 & \frac{1}{2}  & \frac{1}{3} & -\frac{5}{3}   \\[0.2em]
		0  & 0  & 2 & \frac{4}{3}  & -\frac{4}{9} & -\frac{8}{9}   \\[0.2em]
		\hline\rule{0pt}{0.9\normalbaselineskip}
		0  & 0  & 0 & 0 & -\frac{35}{9} & -\frac{52}{9}   \\[0.2em]
		0  & 0  & 0 & 0  & -\frac{110}{9} & \frac{5}{9}   \\[0.2em]
	\end{array}
	\right)\,.
	\end{align}
One can see that anomalous dimensions from 
dim-10 length-5 evanescent operators are all real.

\subsection{Dimension 12,14,16}

The Gram matrices and one-loop dilatation matrices of dim-12,14,16 operators
are too large to be presented here, and we provide them in the ancillary file.
In Section~\ref{theadresults} we show the one-loop dilatation matrices of
two subsectors in dim-12 length-4 and dim-12 length-5.

\section{The number of complex roots in YMS}
\label{root analysis}
This section shows how we calculate the number of complex roots in YMS. Below we first give a short introduction to Sturm's chain and  Sturm's theorem (see \emph{e.g.} \cite{basu2003algorithms} for a more detailed description). Then we describe our analysis for the equation \eqref{thexequation}.

Given a polynomial in one variable, say $f(x)$, its Sturm's chain is the sequence of polynomials constructed as below:
\begin{align}
	&s_0=f(x)\,,\nonumber\\
	&s_1=f'(x)\,,\nonumber\\
	&s_3=-\text{Rem}(s_{1},s_{2})\,,\nonumber\\
	&\dots\nonumber\\
	&s_i=-\text{Rem}(s_{i-2},s_{i-1})\,,\nonumber\\
	&\dots\,,\label{theschain}
\end{align}
where ``Rem$(f,g)$" means the remainder of $f$ divided by $g$ via the polynomial division. We denote the number of sign changes of the chain \eqref{theschain} at $x=x_0$ as $\sigma_{x_0}$. Sturm's theorem states that given an interval $(a,b)$, the number of distinct real roots for the equation $f(x)=0$ is equal to $\Delta(a,b)\equiv\sigma_{a}-\sigma_{b}$~$(a<b)$, provided that neither a nor b is a root. In particular, $\Delta(-\infty,\infty)$ is the number of all distinct real roots.

Let us consider the eigenvalue equation \eqref{thexequation} of YMS in Section~\ref{matter}. The polynomial on the l.h.s. is of degree 10 and with a parameter $N_f$. According to \eqref{theschain}, one can construct  Sturm's chain that includes eleven elements. At $\pm\infty$, the signs of the chain are solely determined by the signs of the coefficients of the leading terms. We denote the leading terms as
\begin{align}
	LD(s_0)=c_{10} x^{10},\ LD(s_1)=c_{9} x^{9},\ \dots,\ LD(s_{10})=c_{0}\,,
	\label{ldsturm}
\end{align}
where $LD$ means the leading term and $c_i$'s are rational functions of $N_f$. 

We first analyze the signs of \{$c_i$\} for within the region $N_f\geq 1$. At $N_f=1$, the signs of $c_i$'s (the arrangement is $\{c_{10},c_9,\dots,c_0\}$)
\begin{align}
	\{+,+,+,+,+,+,+,-,-,-,-\}\,, \label{pnf1}
\end{align}
Go on to consider the signs of \{$c_i$\} within the region $N_f>1$. For each rational function $c_i$, the sign changes when it passes through a root of its numerator or of its denominator, provided that the root has odd multiplicity. By an elementary one-by-one analysis, we find that only the signs of $c_3$ and $c_4$ would change within the region $N_f>1$:
\begin{align}
	1\leq N_f\leq 8,&\quad c_4>0\text{ and }c_3<0\,,\label{theD6}\\
	9\leq N_f\leq 13,&\quad c_4<0\text{ and }c_3<0\,,\\
	14\leq N_f\leq 24,&\quad c_4>0\text{ and }c_3>0\,,\\
	N_f\leq 25,&\quad c_4>0\text{ and }c_3<0\,.\label{theD8}
\end{align}

Having the signs of \{$c_i$\}, it is straightforward to calculate $\Delta(-\infty,\infty)$. It turns out that $\Delta(-\infty,\infty)=8$ for any positive integer $N_f$. According to Sturm's theorem, there are 8 real roots for the eigenvalue equation \eqref{thexequation}. Besides, by numerically calculating the roots of the numerators and denominators of these rational functions, one can verify that there is no positive integer root for each $c_i$. This implies that  Sturm's chain does not terminate and one can get a non-vanishing $c_0$ for all positive integer $N_f$. Therefore the polynomial in \eqref{thexequation} and its derivative have no common factor, implying that \eqref{thexequation} has 10 distinct roots in the complex plane $\mathbb{C}$. Therefore, the eigenvalue equation \eqref{thexequation} has $2$ complex roots for all positive integer $N_f$.


\bibliographystyle{utphys}

\providecommand{\href}[2]{#2}\begingroup\raggedright\endgroup

\end{document}